\documentclass[aps,prd,reprint,nofootinbib,showkeys,showpacs,superscriptaddress,longbibliography,floatfix]{revtex4-2}[12pt]

\pdfoutput=1

\usepackage{amsmath}
\usepackage{amssymb}
\usepackage[american]{babel}
\usepackage{booktabs}
\usepackage{dcolumn}  % Align table columns on decimal point
\usepackage[T1]{fontenc}  % Font encoding that enables acents in the PDF
\usepackage{graphicx}  % Include figure files
\usepackage[colorlinks]{hyperref}  % Add hypertext capabilities
\usepackage[utf8]{inputenc}  % Accept the Western European characters set
\usepackage{multirow}
\usepackage{txfonts}  % Fonts: Times (text) and Txfonts (math)
\usepackage{url}
\usepackage[dvipsnames]{xcolor}
\usepackage{float}
\usepackage{bm}% bold math
\usepackage{lipsum}

\newcolumntype{d}[1]{D{.}{.}{#1}}
\newcolumntype{v}[1]{D{,}{,\ }{#1}}

\makeatletter

\newcommand{\Rmnum}[1]{\expandafter\@slowromancap\romannumeral #1@}
\makeatother

\hypersetup{
	citecolor = cyan,
	linkcolor = red,
	urlcolor = magenta
}

\begin{document}

	\title{Growth of matter perturbations in the extended viscous dark energy models}

	\author{W. J. C. da Silva}
	\email{williamjouse@fisica.ufrn.br}
	\affiliation{Universidade Federal do Rio Grande do Norte, Departamento de F\'{\i}sica, Natal - RN, 59072-970, Brazil}
    
    \author{R. Silva}
	\email{raimundosilva@fisica.ufrn.br}
	\affiliation{Universidade Federal do Rio Grande do Norte, Departamento de F\'{\i}sica, Natal - RN, 59072-970, Brazil}
	\affiliation{Universidade do Estado do Rio Grande do Norte, Departamento de F\'{\i}sica, Mossor\'o - RN, 59610-210, Brazil}
	
	\date{\today}

	\begin{abstract}
	In this work, we study the extended viscous dark energy models in the context of matter perturbations. To do this, we assume an alternative interpretation of the flat Friedmann-Lemaître-Robertson-Walker Universe, through the nonadditive entropy and the viscous dark energy. We implement the relativistic equations to obtain  the growth of matter fluctuations for a smooth version of dark energy.  As result, we show that the matter density contrast evolves similarly to the $\Lambda$CDM model in high redshift; in late time, it is slightly different from the standard model.  Using the latest geometrical and growth rate observational data, we carry out a Bayesian analysis to constrain parameters and compare models. We see that our viscous models are compatible with cosmological probes, and the $\Lambda$CDM recovered with a $1\sigma$ confidence level. The viscous dark energy models relieve the tension of $H_0$ in $2 \sim 3 \sigma$. Yet, by involving the $\sigma_8$ tension, some models can alleviate it. In the model selection framework, the data discards the extended viscous dark energy models.
	\end{abstract}

	\maketitle
	
	\section{Introduction}\label{sec:intro}
	
	Although the current cosmological observations corroborate with both the accelerated expansion of the Universe  and the standard $\Lambda$CDM model  \cite{Riess1998,Perlmutter1999,Eisenstein2005,Percival2010,Reid2012,Ade2016,Aghanim2020,Weinberg2013}, some theoretical and observational problems have been challenging this scenario. From the theoretical standpoint, fine-tuning and cosmic coincidence are the shortcomings of the model \cite{Weinberg1989,Zlatev1999,Sahni2000,Padmanabhan2003,Velten2014}. More recently, significant tensions were associated with the measurements from the Hubble constant $H_0$ and the amount of matter content $\sigma_8$. The inconsistency between the local measurement of the $H_0$, reported as $H_0=74.03\pm 1.42$ km/s/Mpc in \cite{Riess2019}, and its estimation through the Planck cosmic microwave background (CMB), with the value $H_0=67.36\pm 0.54$ km/s/Mpc (a discrepancy of $\sim 4.4 \sigma$) \cite{Aghanim2020}, certainly emerges as the greatest challenge of the modern cosmology (see e.g. \cite{Freedman2017} and references therein). Another important issue is the tension on the measurements of $\sigma_8$, which is based on the large scale structure observations, and from Planck CMB results \cite{Battye2015}.
    
    These issues open alternative approaches beyond the $\Lambda$CDM model; such ones consider either modifications of general relativity \cite{Clifton2012} or new description for dark energy (DE) \cite{Huterer2017} or different theories of dark matter (DM) \cite{Vattis2019}. There are many approaches in order to investigate the DE, e.g., the dynamical DE, phantom DE, quintessence, Chaplygin gas, holographic DE, the interaction between DE and DM, phenomenological emergent DE, and early DE (see Refs. \cite{Huterer2017,Li2004,Copeland2006,Li2019,Rezaei2020,Pan2019,Hernandez-Almada2020,Poulin2019,Yang2020,DiValentino2020}). Among several models to investigate the DE, the thermodynamic approach has been widely investigated (see,  e.g.,  Refs. \cite{Lima1992,Lima2004,Izquierdo2006,Pereira2008,Lima2010,Silva2011,Normann2016,Gonzalez2018,daSilva2020}, and the references therein). Also, an interesting proposal considered an exotic fluid with bulk viscosity. Models assuming this possibility have also been studied in the context of the interaction of DE and DM \cite{Velten2012,Avelino2013,Velten2014,Floerchinger2015,Sasidharan2018,HernndezAlmada2020}. Moreover, those models could resolve the tension across different probes as the values of the Hubble constant, and the tension associated with values matter fluctuation amplitude \cite{Anand2017}. DE models with bulk viscosity have obtained good results by a description based on fluids \cite{Wang2017}, modified general gravity \cite{Mostaghel2017, Mostaghel2018}, and a scalar fluid framework \cite{Gagnon2011}. 
	 
	 Nonextensive statistical mechanics is widely used in the complex systems viewpoint \cite{Tsallis-web,Gell-Mann-book2004}. The core of this framework is associated with the parametrization of the entropy formula, so-called the Tsallis entropy. Such an expression depends on a free parameter $q$, and provides the Boltzmann-Gibbs (BG) entropy in
     the additive limit $q=1$. Recently, many investigations using Tsallis entropy have been used to address issues in cosmology. The first connections between Tsallis framework and cosmology proposed the generalized forms of the first and second laws of thermodynamics in the context of cosmic blackbody radiation in a Robertson-Walker model of the Universe \cite{Hamity1996} and assuming an early Universe scenario, used data of primordial helium abundance in order to investigate nonextensive effects \cite{Torres1997}. More recently, applications have been proposed as regards many issues, e.g., a generalized black hole entropy \cite{Tsallis2013}, viscous dark matter \cite{Gimenes2018} and the modified Friedmann equations using the Verlinde theory \cite{Abreu2013, Nunes2016,daSilva2019, Silva2019}. In the context of DE, this theory studied the dynamical DE, the interaction between DE and DM \cite{Barboza2015}, viscous dark energy \cite{Silva2019} and  holographic dark energy \cite{Tavayef2018, Saridakis2018, DAgostino2019}.\footnote{For an updated list of the papers; see \cite{Tsallis-web}.}

    From the perturbative perspective, on the other hand, the galaxies and galaxy clusters that we observe today have been formed from the initial fluctuations at the inflation era \cite{Linde1990,Peebles1993}. In the cosmic evolution, gravity can amplify the amplitude of matter fluctuations, in particular at the matter-dominated epoch. DE not only accelerates the Universe but also changes the growth rate of matter perturbations and, consequently, the formation periods of large scale structure of the Universe. Furthermore, considering the background geometrical data, the information coming from the structure formation provides knowledge about the nature of DE. Recently, some studies have investigated the effect of DE models on the clustering of matter considering the structure formation measured through redshift-space distortions (RSD) \cite{Abramo2007, Tsujikawa2013,Batista2014,Mehrabi2015,Rezaei2017,Mehrabi2018}. In this regard,  the effect of bulk viscosity in perturbative level has also been investigated in the DM and DE contexts \cite{Piattella2011,Velten2014b,Anand2017,Blas2015,Thomas2016}. Taking into account that RSD data spans a wide range of redshift ($z \approx 1.52$), it is worth exploring its constraint power to distinguish between DE models.

	Recently, the extended $\Lambda$CDM model and viscous DE have been investigated in the background level \cite{Silva2019}.  As a natural extension, the investigation of this scenario in the context perturbative and its mitigation of the cosmic tension are the main goals of this paper. Thus, we study this scenario in the context of the growth of matter perturbations. To this end, we will implement the perturbative relativistic equations to obtain the growth of matter fluctuations in the presence of the extended viscous DE model in order to study its contributions to the rate of structure formation. To constrain cosmological parameters, we will confront it with a combination of different observational datasets using Bayesian inference as a most useful statistical analysis in modern cosmology.

	This work is organized as follows: in Sect. \ref{sec:phenomenology}, we resume the extended viscous dark energy. This section is divided into two subsections \ref{subsec:background} and \ref{subsec:growth}. In Sect. \ref{subsec:background}, we outline the principal equations that describe the extended viscous dark energy. Further, in \ref{subsec:growth}, we show the equations that govern the perturbation level in the presence of this viscous model. In Sect. \ref{sec:data-method}, we describe the cosmological data as well as the methodology used in this work. In Sect. \ref{sec:results}, we present the main results of our work, parameter constraints, Bayesian model selection and cosmic age crisis. Finally, in Sect. \ref{sec:conclusions}, we conclude this work.
	
	\section{Phenomenology of the extended viscous dark energy Model}\label{sec:phenomenology}
	
	\subsection{Background level}\label{subsec:background}
	
    Let us recall the phenomenology of the extended viscous dark energy models. The cosmological model under study in this work consists of a flat FLRW Universe, which contains dark energy as non-perfect fluid, together with cold dark matter, baryon, and radiation. In this description, the total energy-momentum tensor can be expressed as
	
	\begin{equation}\label{eq:energy-momentum1}
	T^{\mu\nu}_{\rm Total} = T^{\mu\nu} + \Delta T^{\mu\nu},
	\end{equation}
	where $T^{\mu\nu}$ is the energy-momentum tensor which represents the perfect fluid, and $\Delta T^{\mu\nu}$ is a tiny perturbation that corresponds all dissipative processes (heat flux, anisotropic-stress, and bulk viscosity). In a homogeneous and isotropic Universe, the only dissipative process allowed is the bulk viscosity. As is widely known, bulk viscosity plays an important role in the Universe dynamics at the background level because it satisfies the cosmological principle \cite{Weinberg1971}. The first approach to studying the relativistic bulk viscosity process is based on the Eckart theory, and despite the causality problem, it is widely used due to its simplicity. In this scenario, the bulk viscosity pressure is given by \cite{Eckart1940}
	
	\begin{equation}\label{eq:bulk-viscosity}
	\Delta T^{\mu\nu} = \Pi h^{\mu\nu},
	\end{equation}
	where $h^{\mu\nu} = g^{\mu\nu} + u^{\mu}u^{\nu}$ is the projector onto the local rest space of the hydrodynamics four-velocity $u_{\mu}$ and $g_{\mu\nu}$ corresponds to the FLRW metric. The bulk viscosity pressure $\Pi$ depends on time and can be written in terms of the Hubble parameter. As we are assuming a FLRW Universe, the bulk viscosity coefficient is given by $\Pi = -3 \xi H$. Even though this formalism has been widely used to describe inflation, late-time acceleration, background and perturbative levels	\cite{Gron1990,Gagnon2011,Velten2012,Bamba2016,Anand2017,Wang2017,Mostaghel2017,Mostaghel2018,Barbosa2017,Brevik2017,Sasidharan2018,HernndezAlmada2020}, there is a fundamental difficulty which is related with its non-causal behavior \cite{Israel1976,Israel1979,Maartens1996} (see also Ref. \cite{Silva2019} and references therein)

    Now, by choosing a reference frame which the hydrodynamics four-velocity $u_{\mu}$ is unitary $u_{\mu}u^{\mu} = 1$, and considering Eqs. (\ref{eq:energy-momentum1}) and (\ref{eq:bulk-viscosity}), we obtain
	
	\begin{equation}\label{eq:energy-momentum}
	T_{\mu\nu} = (\rho + \tilde{p})u_{\mu}u_{\nu} + \tilde{p}g_{\mu\nu},
	\end{equation}
	where $\rho$ is the energy density, and $\tilde{p} = p + \Pi$ is the effective pressure which consists of a sum of two terms, equilibrium pressure $p$ and bulk viscosity pressure $\Pi$. Hence, the effective pressure $\tilde{p}$ includes the viscous term $\Pi$, which would satisfy the transportation equation and may be extended to more general forms that contain derivatives of the Hubble parameter and energy density $\xi = \xi(z, \rho, H, \dot{H})$ \cite{Odintsov2020}. From the conservation of energy, $\nabla_{\mu}T^{\mu}_{\nu} = 0$, we obtain
	
	\begin{equation}\label{eq:conservation-equation}
	\dot{\rho} + 3H(\rho + p) - 9H^2\xi = 0.
	\end{equation}
	This relation is the energy conservation equation for any viscous fluid. The bulk viscosity coefficient's behavior $\xi$ must be considered for the description to be complete. Here, we will assume several forms of the bulk viscosity coefficient to describe the viscous dark energy.
	
	It was demonstrated in Refs. \cite{Abreu2013,Nunes2016,Barboza2015,Silva2019} that one modification in the dynamics of the FLRW Universe in Tsallis statistics can be achieve by making the prescription $G \rightarrow G_q =  \frac{5-3q}{2}G$ in the standard field equations, with $q$ being the nonadditive parameter. Then, we want to study the viscous dark energy in the context of the extended Friedmann equations
	
	\begin{equation}\label{eq:extended-friedmann}
	H^{2} = \frac{8\pi G}{3}\rho\left(\frac{5 - 3q}{2}\right) - \frac{k}{a^2},
	\end{equation} 
	
	\begin{equation}\label{eq:extended-friedmann2}
	\frac{\ddot{a}}{a} = -\frac{4\pi}{3}\left(\frac{5 - 3q}{2}\right)G(\rho +3p),
	\end{equation}
	where $H = \frac{\dot{a}}{a}$ is the Hubble parameter, $\rho$ is the total energy density, $p$ is the pressure of the perfect fluid and $k$ represents the spatial curvature. We assume the flat Universe, $k = 0$.
	
	Also, we will describe the viscous dark energy in the context of the extended Friedmann Eqs. (\ref{eq:extended-friedmann}) and (\ref{eq:extended-friedmann2}). The main contributions to the total momentum-energy tensor Eq. (\ref{eq:energy-momentum}) are radiation, baryons, cold dark matter, which have the usual properties of perfect fluids, and viscous dark energy. As each fluid is individually conserved, from Eqs. (\ref{eq:conservation-equation}) and (\ref{eq:extended-friedmann}) we obtain the following conservation equations
	
	\begin{subequations}
		\label{eq:continuous-equations}
		\begin{eqnarray}
		&H^{2} = \frac{8\pi G}{3}\left(\frac{5 - 3q}{2}\right)(\rho_{\rm r} + \rho_{\rm m}+\rho_{\rm de}),\label{eq:friedmann-equation} \\
		&\dot{\rho}_{\rm r}+4H\rho_{\rm r}=0, \label{conservation-radiation} \\
		&\dot{\rho}_{\rm m}+3H\rho_{\rm m}=0, \label{conservation-m} \\
		&\dot{\rho}_{\rm de}+3H(1+w_{\rm de})\rho_{\rm de}= 9H^2\xi,  \label{conservation-dark}
		\end{eqnarray}
	\end{subequations}
	where $\rho_{\rm r}$, $\rho_{\rm m}$, $\rho_{\rm de}$ are the radiation, matter (baryon + cold dark matter) and viscous dark energy densities, respectively. The choice of bulk viscosity coefficient provides distinct extended viscous models. It is worth noting that for all models, the dark energy density $\rho_{\rm de}(z)$ is a dynamical function due to its viscosity term. In general, we can consider higher order corrections in energy-momentum tensor, but it was demonstrated that these terms have no significant consequence on the dynamical acceleration \cite{Hiscock1991}. Then, we have considered the three functional forms of $\xi$ in our analysis, namely: \textit{i)} bulk viscosity proportional to the Hubble parameter, $\xi = \eta H$; \textit{ii)} bulk viscosity proportional to energy density and inversely proportional to Hubble parameter, $\xi =\eta\sqrt{\rho_{\rm de}}/H$; \textit{iii)} thermodynamic function state, $\xi = \eta\rho^{\nu}_{\rm de}$. Let us discuss a little bit more about these models 
	
	\subsubsection{Model 1}
	
	The first model here considered was studied in Ref. \cite{Wang2017}. The functional form of the bulk viscosity is proportional to the Hubble parameter, which means that $\xi$ is proportional to the total energy density's square root. Based on the Friedmann Eq. (\ref{eq:friedmann-equation}), we can consider the bulk viscosity to be a function of all the other cosmological components in the Universe. The bulk viscosity is given by

	\begin{equation}\label{eq:bulk-1}
	\xi = \eta_{0} H,
	\end{equation}
	where $\eta_{0}$ is the current value of the bulk viscosity, then from Eqs. (\ref{eq:continuous-equations}) with $w_{\rm de} = - 1$, we can express the Hubble parameter as

	\begin{equation}\label{eq:hubble1}
	\begin{aligned}
	\frac{H^{2}}{H_{0}^2}  =  \left(\frac{5 - 3q}{2}\right) \Bigg[\Omega_{\text{r}}(1 + z)^{4} & + \frac{\Omega_{\text{m}}}{1 + \eta}(1 + z)^3 \\ & + \Bigg(1 - \frac{\Omega_{\text{m}}}{1 +  \eta}\Bigg)(1 + z)^{-3\eta}\Bigg],
	\end{aligned}	
	\end{equation}
	where $H_{0}$ is the Hubble constant, $q$ is the nonaddictive parameter, $\eta$ is the bulk viscosity constant and $z$ is the redshift. The dimensionless bulk viscosity constant is defined by

	\begin{equation}
	\eta = \frac{8\pi G \eta_{0}}{H_0}.
	\end{equation}

	\subsubsection{Model 2}
	
	The second model was proposed in the Refs. \cite{Mostaghel2017,Mostaghel2018}. The functional form of the bulk viscosity is a ratio between dark energy density and expansion rate given by
	
	\begin{equation}\label{eq:bulk-viscosity2}
	\xi= \eta_0\frac{\sqrt{\rho_{\text{de}}}}{H},
	\end{equation}
	where $\eta_0$ is the value of bulk viscosity computed in the present-day and $\rho_{\text{de}}$ is the dark energy density. By assuming this form of the bulk viscosity, we can note that contribution of the viscous dark energy in the early Universe is slight. Otherwise, at late-time, this contribution is relevant and the viscous dark energy dominated the Universe evolution \cite{Mostaghel2017,Mostaghel2018}.
	
	Next, considering Eqs. (\ref{eq:continuous-equations}) with $w_{\rm de} = - 1$ and Eq. (\ref{eq:bulk-viscosity2}), we obtain the Hubble parameter for model 2
	
	\begin{equation}\label{eq:hubble2}
	\begin{aligned}
	\frac{H^{2}}{H_{0}^2} = & \left(\frac{5 - 3q}{2}\right) \Bigg[\Omega_{\text{r}}(1 + z)^{4} +  \Omega_{\text{m}}(1+z)^{3} \\ & + \Omega_{\text{de}}\Bigg(1 - \frac{9\eta}{2\sqrt{\Omega_{\text{de}}}}\ln (1 + z)\Bigg)^2\Bigg],
	\end{aligned}
	\end{equation}
	where $\Omega_{\text{de}} = \frac{2}{5-3q} - \Omega_{\text{m}} - \Omega_{\text{r}}$. The dimensionless bulk viscosity parameter $\eta$ is defined by
	
	\begin{equation}
	\eta = \sqrt{\frac{8\pi G}{3H^2_0}}\eta_{0}.
	\end{equation}
	
	\subsubsection{Model 3}
	
	Model 3 was studied in the context of viscous dark matter \cite{Velten2012}. The bulk viscosity is assumed as a thermodynamic state of the energy density of the respective fluid. In the case of dark energy, the bulk viscosity is given by

	\begin{equation}\label{xi-model3}
	\eta = \eta_{0}\Bigg(\frac{\rho_{\text{de}}}{\rho_{\text{de}0}}\Bigg)^{\alpha},
	\end{equation}
	where $\eta_{0}$, $\rho_{\text{de}0}$ are the current values for bulk viscosity and density of the viscous dark energy and $\alpha$ is constant. In the case of viscous dark matter, the functional form this bulk viscosity can alleviate the integrated Sachs-Wolfe effect \cite{Velten2012}. By fixing $\alpha = 0$, then from Eq. (\ref{conservation-dark}) with $w_{\rm de} = -1$, we obtain

	\begin{equation}\label{eq:ode}
	\begin{aligned}
	\frac{d \Omega_{\text{de}}}{dz} = & -  \frac{\eta}{1 + z}\Bigg\{\left(\frac{5 - 3q}{2}\right) \Big[\Omega_{\text{r}}(1 + z)^{4} \\ & + \Omega_{\text{m}}(1 + z)^{3} + \Omega_{\text{de}}(z)\Big]\Bigg\}^{1/2},
	\end{aligned}	
	\end{equation}
	with initial condition $\Omega_{\text{de0}} = \frac{2}{5-3q} - \Omega_{\text{m}} - \Omega_{\text{r}}$.

	From the Friedmann equation (\ref{eq:friedmann-equation}), we obtain the Hubble expansion rate $H$ in terms of the energy densities, 

	\begin{equation}\label{eq:hubble3}
	\frac{H^{2}}{H_{0}^{2}} =  \left(\frac{5 - 3q}{2}\right) \Big[\Omega_{\text{r}}(1 + z)^{4} + \Omega_{\text{m}}(1 + z)^{3} + \Omega_{\text{de}}(z)\Big],
	\end{equation}
	where $\Omega_{\text{de}}(z)$ is given by the solution of Eq.(\ref{eq:ode}). The dimensionless bulk viscosity coefficient is defined by

	\begin{equation}
	\eta = \frac{24 \pi G \eta_{0}}{H_{0}}.
	\end{equation}

    The standard $\Lambda$CDM model is recovered in the limit $\eta \rightarrow 0$ and $q \rightarrow 1$ for all models. In the next section, we will study these models in the context of the matter perturbations. 
	
	\subsection{Growth of perturbations}\label{subsec:growth}
	
	\begin{figure*}
		\centering
		\includegraphics[width=8.5cm]{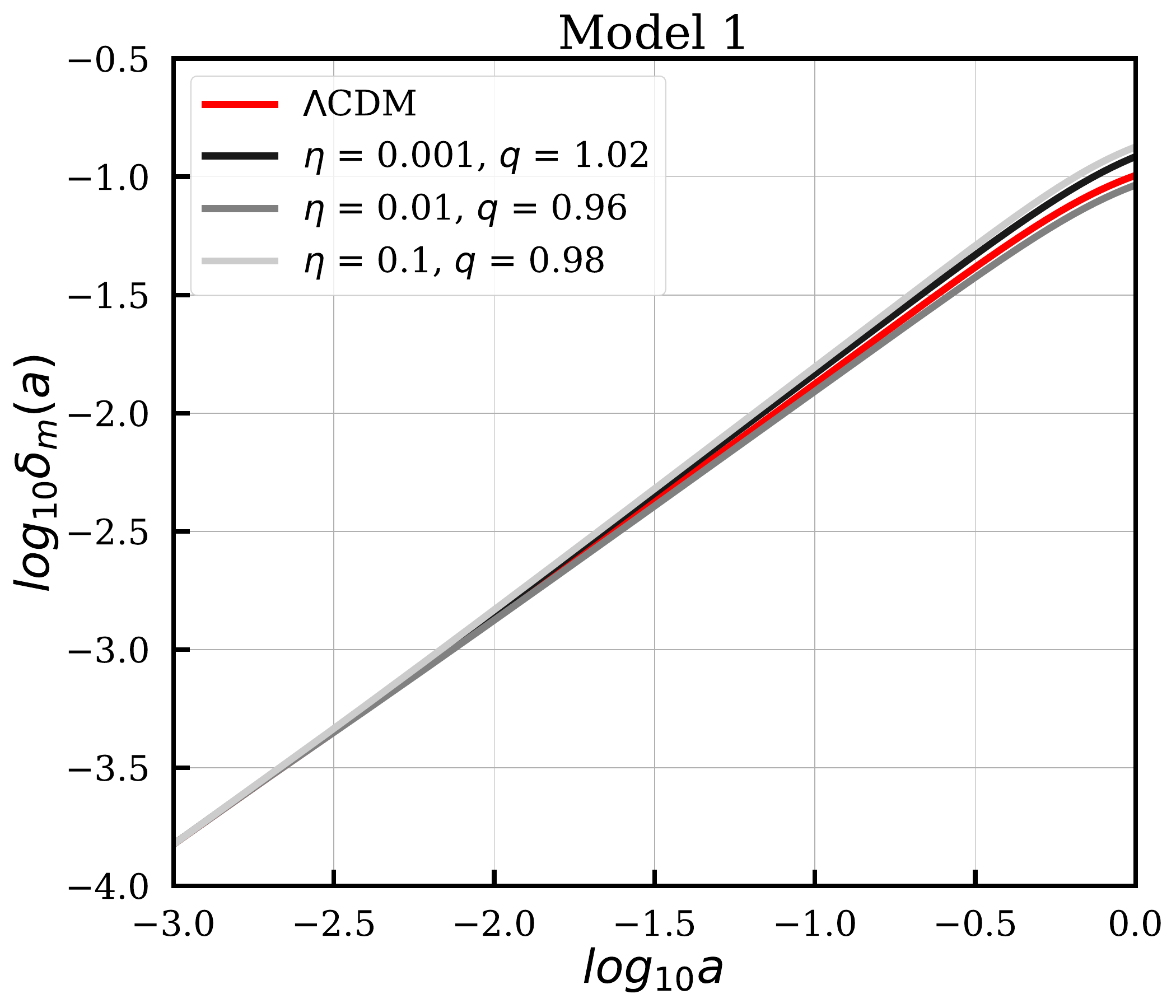}
		\includegraphics[width=8.5cm]{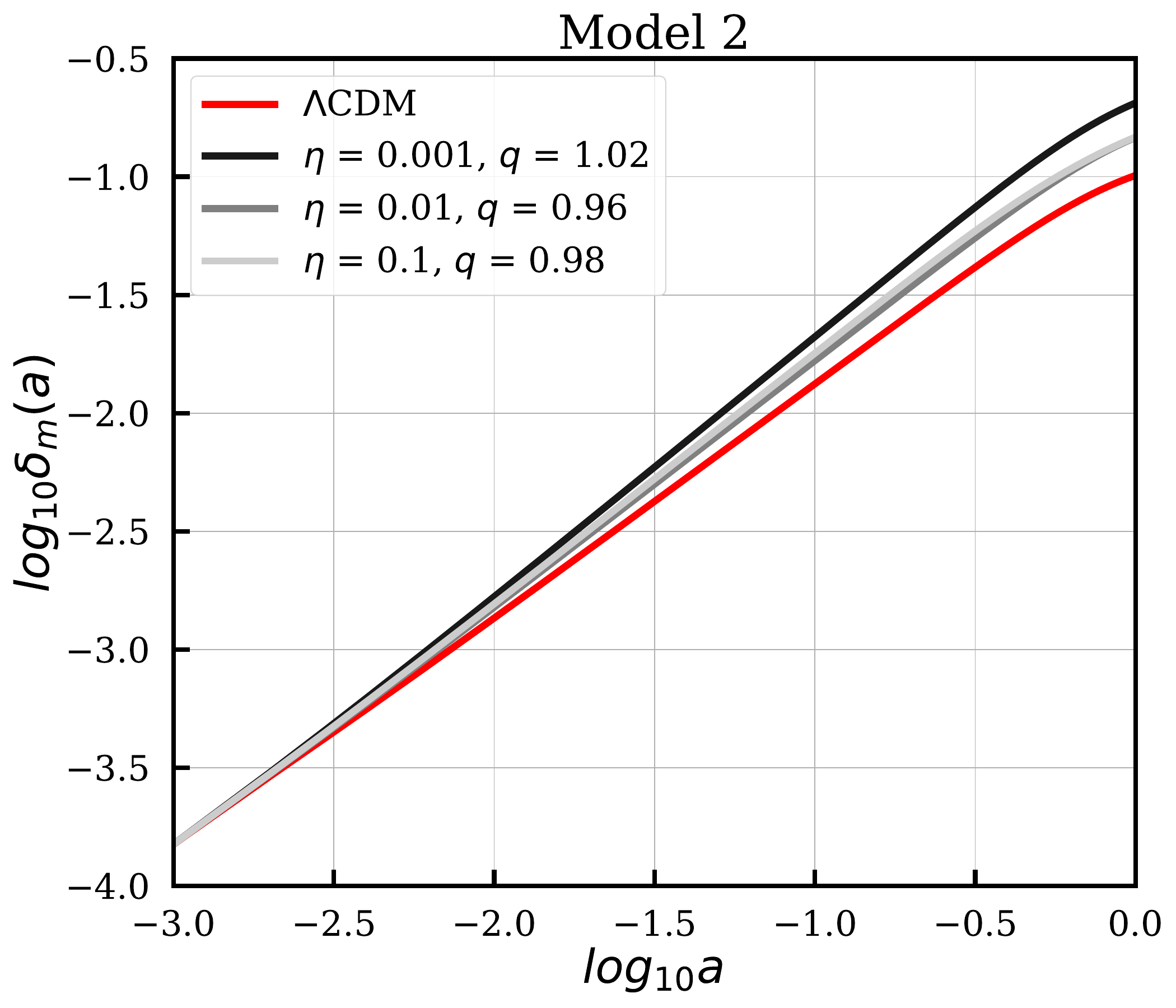}
		\includegraphics[width=8.5cm]{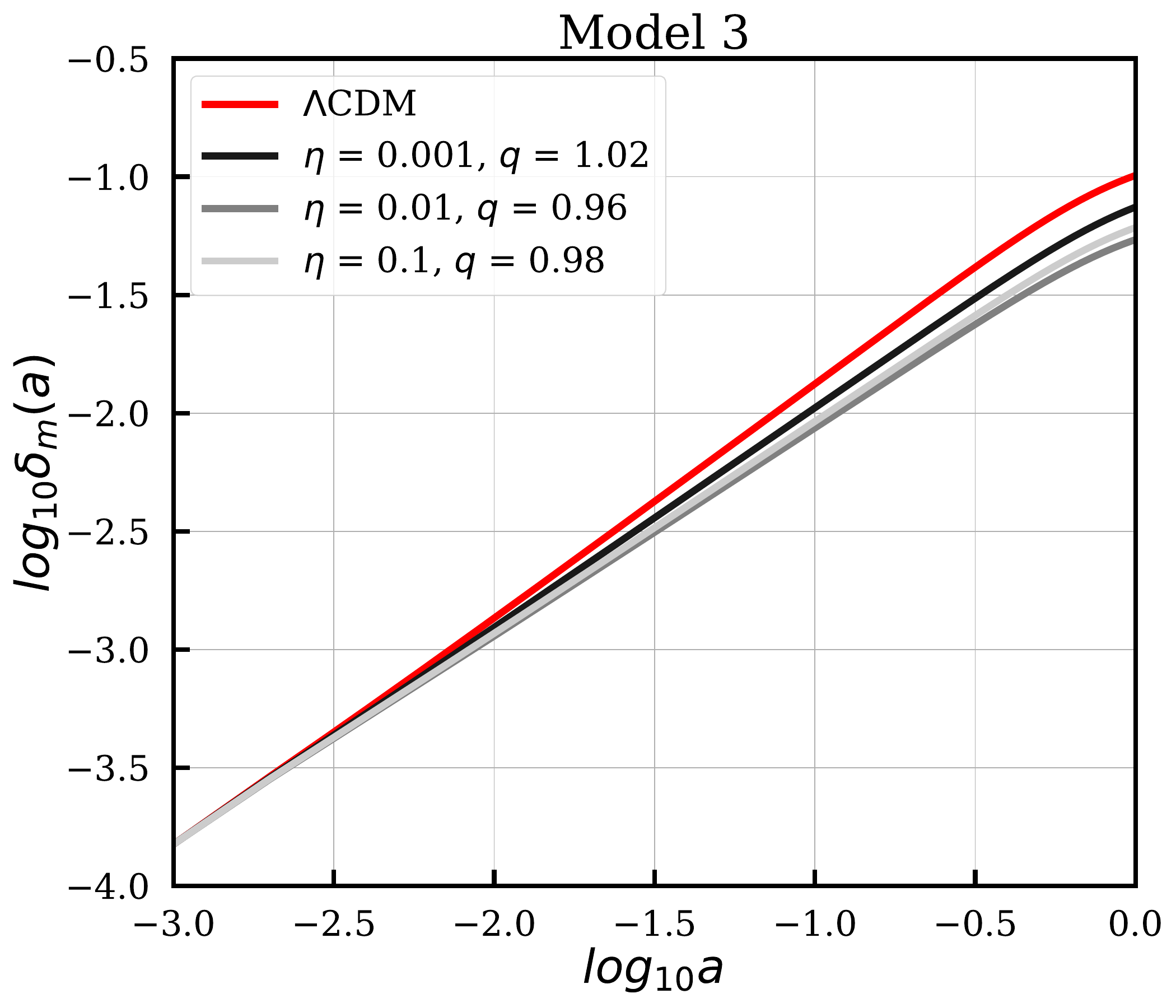}
		\includegraphics[width=8.5cm]{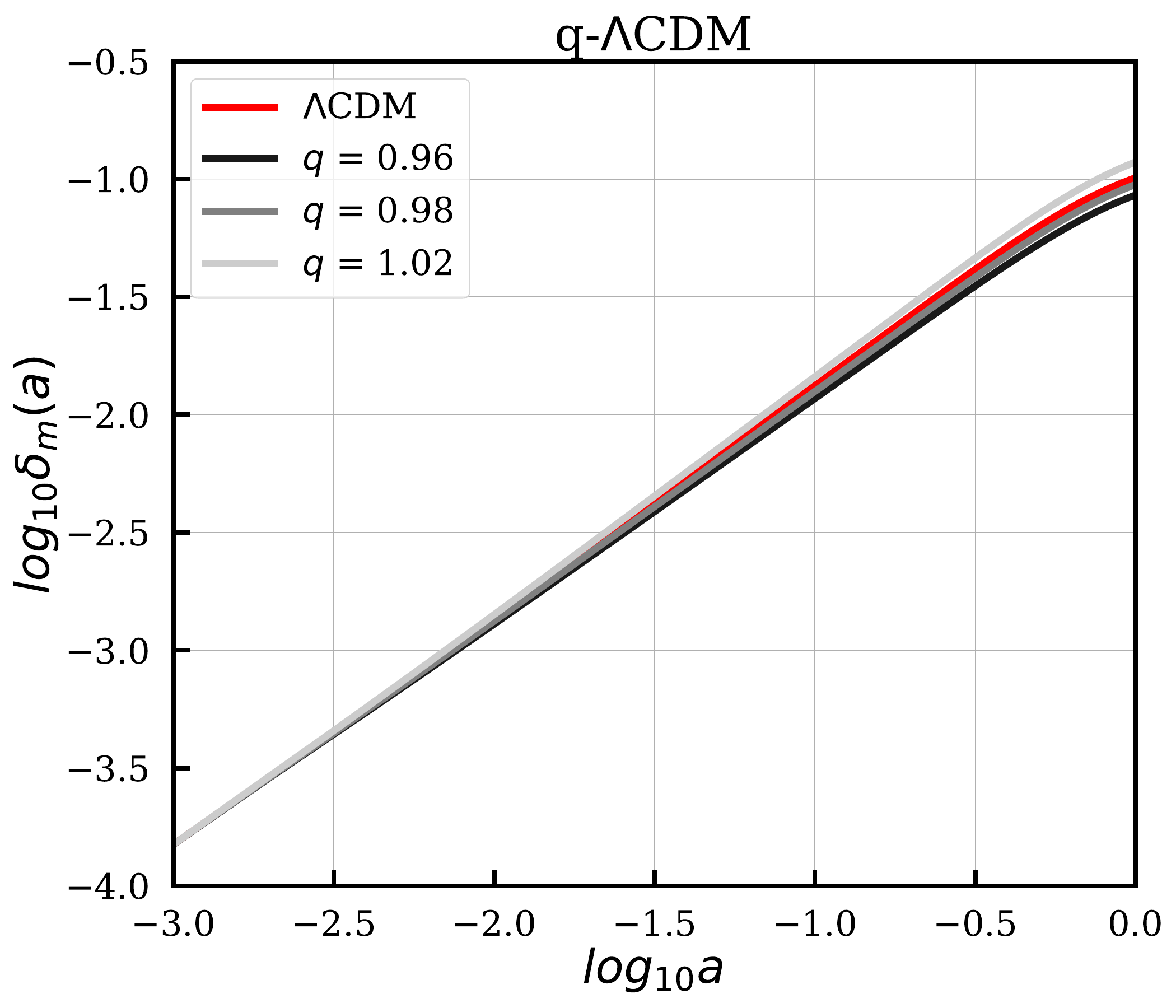}
		\caption{\label{fig:delta_ms} The evolution of perturbations of matter as a function of scale factor for different parameter combinations for Model $1$, $2$, $3$ and q-$\Lambda$CDM.}
	\end{figure*}
	
	The standard theory of cosmic structure formation assumes that the present abundant structure of the Universe developed through gravitational amplification of small matter density perturbations generated in its early history. This section briefly reviews the main mathematical form of the linear perturbation theory within the framework of extended viscous dark energy cosmologies.

	\begin{figure*}
		\centering
		\includegraphics[width=8.5cm]{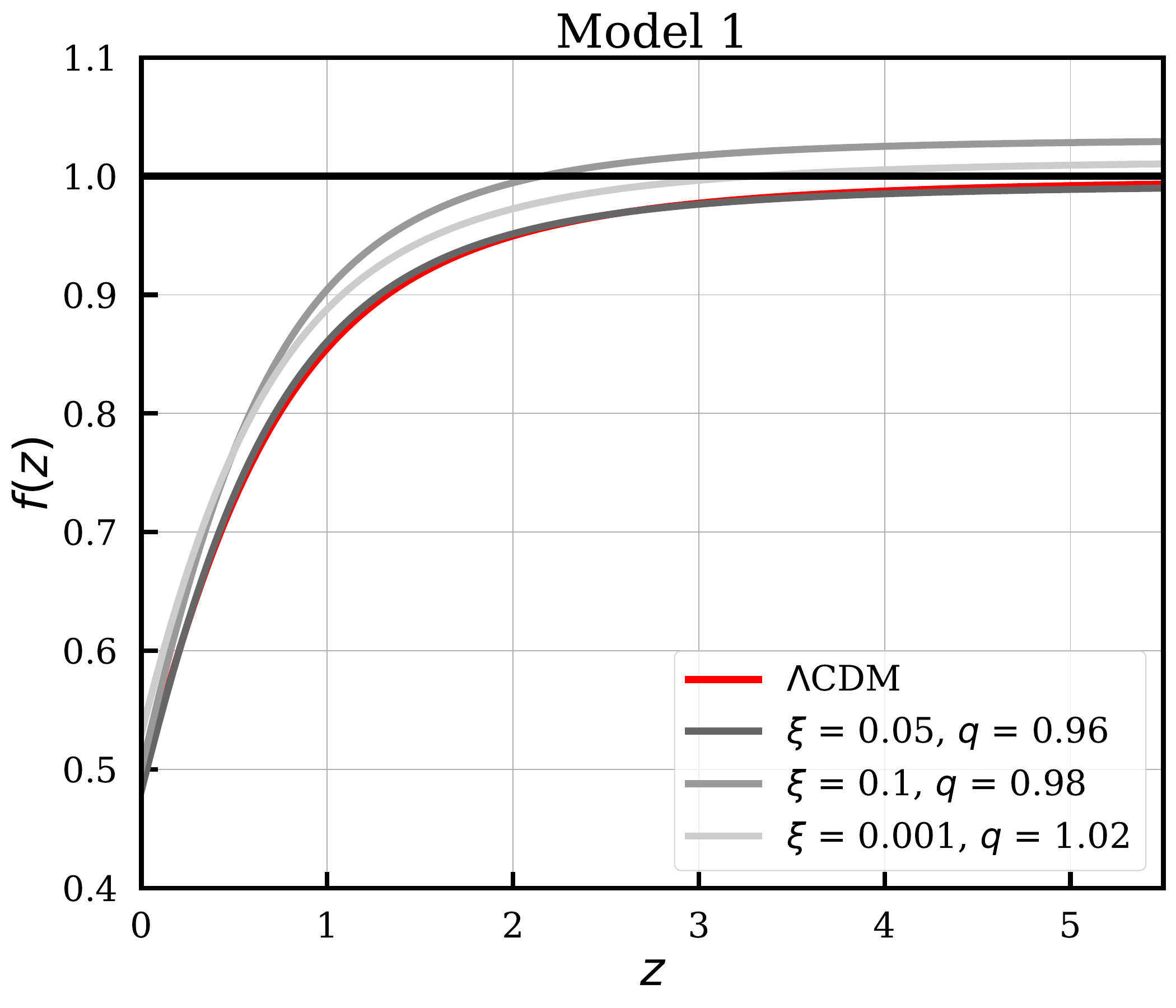}
		\includegraphics[width=8.5cm]{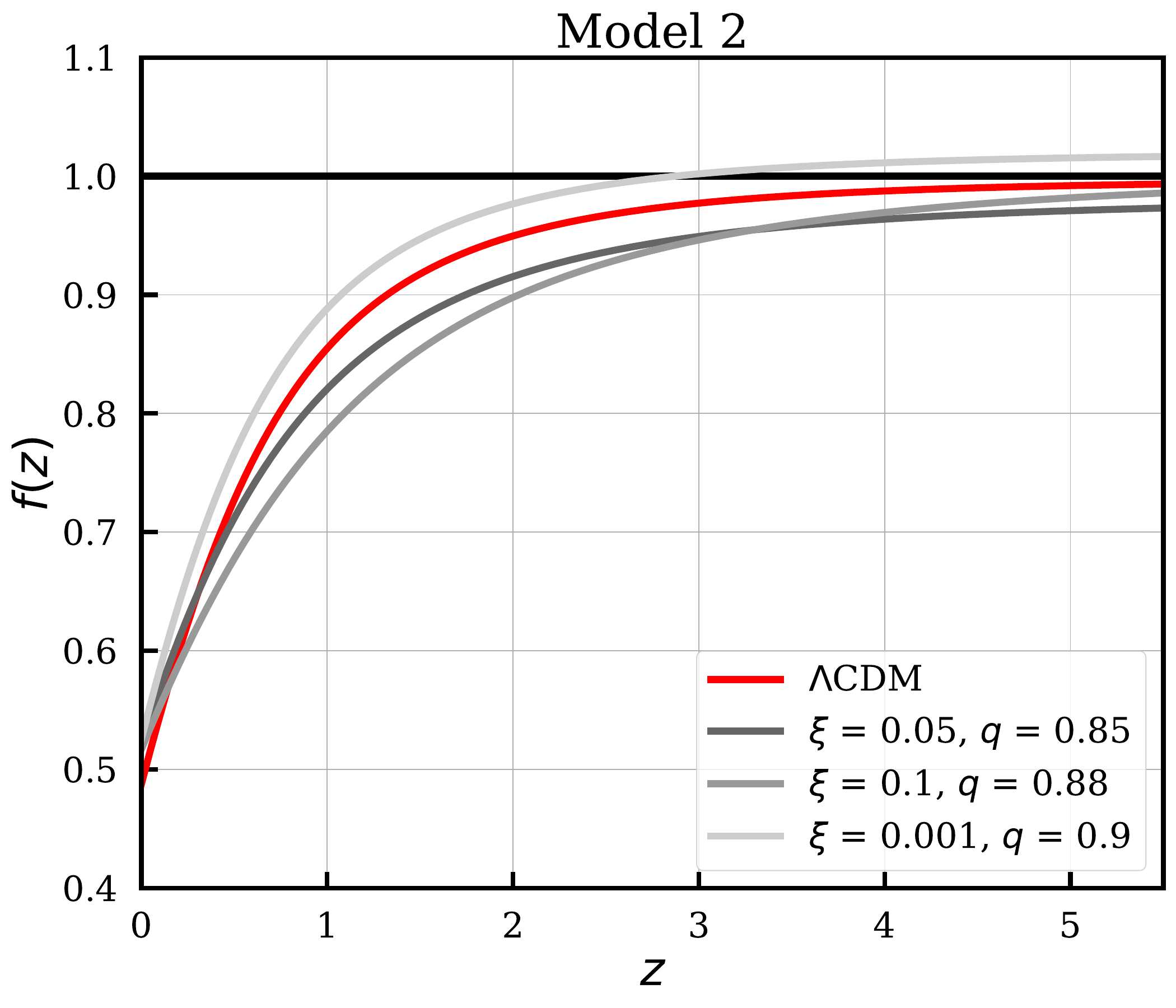}
		\includegraphics[width=8.5cm]{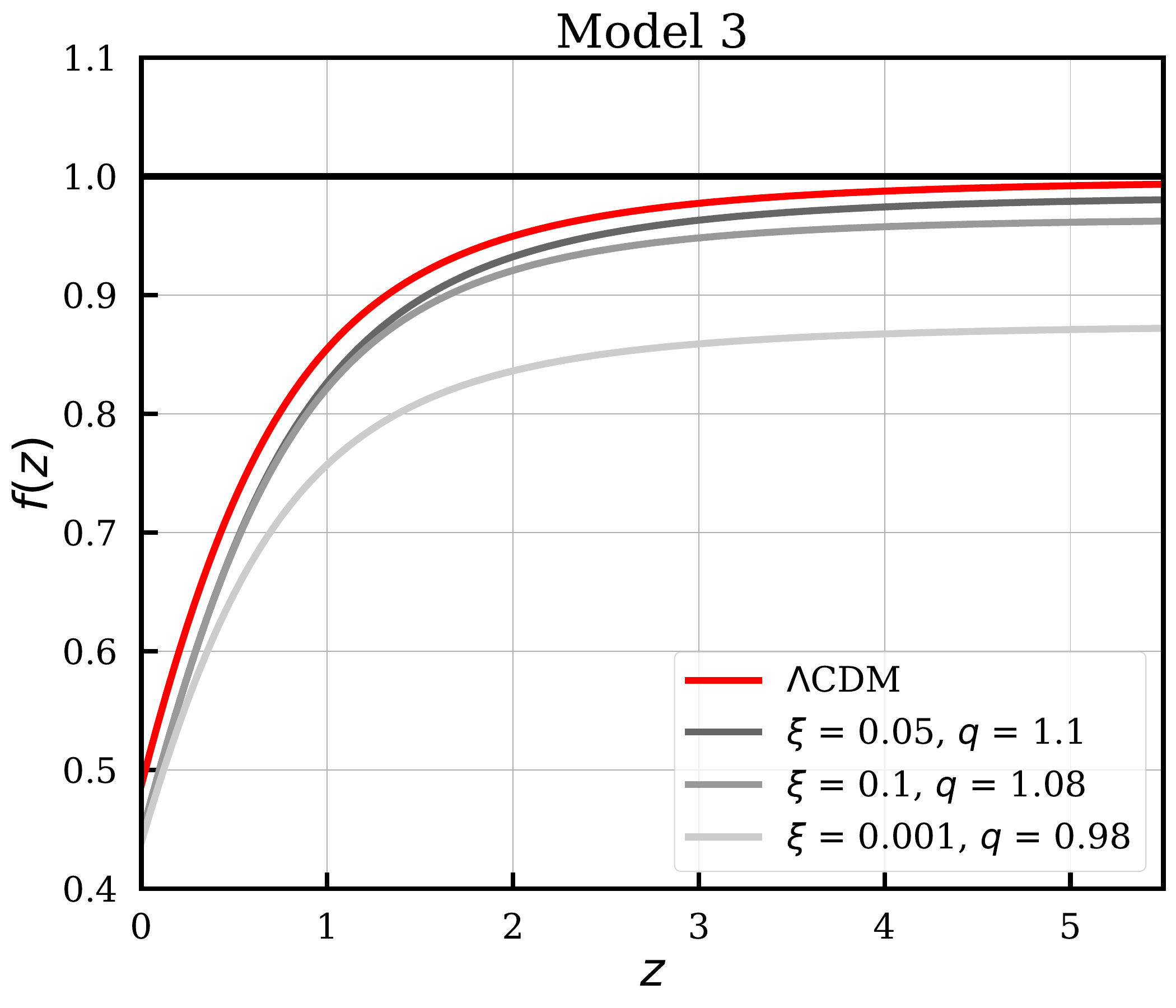}
		\includegraphics[width=8.5cm]{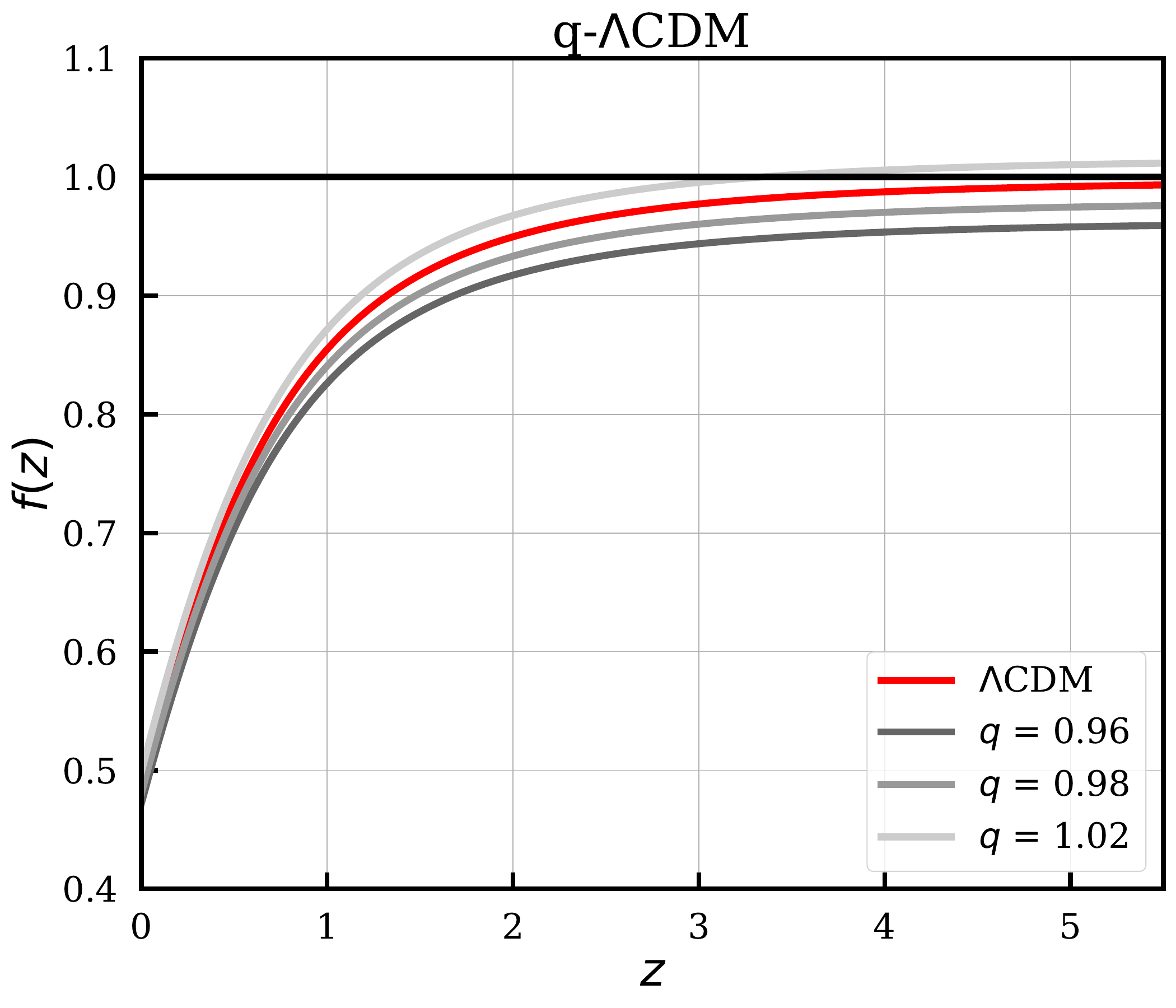}
		\caption{\label{fig:DMs}The evolution of linear growth function as a function of scale factor for different parameter combinations for Model $1$, $2$, $3$, and q-$\Lambda$CDM.}
	\end{figure*}
	
	In order to study the evolution of matter perturbations in the presence of extended viscous dark energy cosmologies, we hold Einstein equations for tiny inhomogeneities as 
	\begin{equation}
	\delta G^{\mu}_{\nu} = 8 \pi G \delta T^{\mu}_\nu.
	\end{equation}
	
	We consider scalar fluctuations of the FLRW metric in the conformal Newtonian gauge \cite{Bardeen1980,Kodama1984,Mukhanov1992}
	
	\begin{equation}
	ds^2 =  a^2(t)[(1+2\phi)d\eta^2 -(1-2\phi)\delta_{ij}dx^{i}dx^{j}],
	\end{equation}
	where $\eta$ is the conformal time, $a$ is the scale factor and $\phi$ is the Bardeen potential. The Einstein equations in the perturbed FLRW metric read \cite{Ma1995}
	
	\begin{subequations}
		\label{eq:einstein-equations}
		\begin{eqnarray}
		k^2\phi - 3\mathcal{H}\left(\phi' + \mathcal{H}\phi\right) &=& 4\pi Ga^2\delta\rho, \label{eq:first-order-En-eq1}\\
		k^2\left(\phi' + \mathcal{H}\phi\right) + \left(\mathcal{H}^2 - \mathcal{H}'\right)\Theta &=& 0, \label{eq:first-order-En-eq2}\\
		\phi'' + 3\mathcal{H}\phi' + \left(\mathcal{H}^2 + 2\mathcal{H}'\right)\phi &=& 4\pi Ga^2\delta p \label{eq:first-order-En-eq3},
		\end{eqnarray}
	\end{subequations}
	where $\mathcal{H} = aH$ is the Hubble parameter as a function of the conformal time, and the prime denotes a derivative with respect to the conformal time. In these equations, $\delta\rho$ is the perturbation of density of the fluid, and $\Theta = \partial_i v^i$ is the divergence of velocity. Observe that in Eqs. (\ref{eq:einstein-equations}) we consider the general case in which both dark matter and viscous dark energy have been perturbed. It is possible to derive a set of differential equations representing the perturbations in the extended viscous dark energy component. Nonetheless, we follow Ref. \cite{Mostaghel2018} and adopt suitable initial conditions to suppress the dark energy perturbations, and finally, the effect of instability is resolved.\footnote{In this regard, the three models discussed have presented the contrast of the viscous dark energy, being either null or negative. It results in a cosmological constant or unphysical behavior, respectively.} So, we restrict our analysis to the scenario in which we only have dark matter clustering.

	The conservation of energy-momentum is a consequence of the Einstein equations. Then the perturbed part of energy-momentum conservation equations in Fourier space reads

	\begin{subequations}
	\label{eq:continuity-equations}
	\begin{eqnarray}
	\delta' &=& -(1 + w)(\Theta - 3\phi') - 3\mathcal{H}\left(\frac{\delta p}{\delta \rho} - w\right)\delta,  \label{eq:continuity1}\\
	\theta' &=& -\mathcal{H}(1 - 3w)\Theta - \frac{w'}{1 + w}\Theta + \frac{\delta p/\delta \rho}{1 + w}k^2\delta + k^2\phi \label{eq:continuity2}.
	\end{eqnarray}
	\end{subequations}
    These equations are general. They can describe the evolution of cold dark matter and dark energy. In sub-horizon scales ($\mathcal{H}^2\ll k^2$) and in the matter domination epoch ($\phi\approx cte$), we combine the Eqs. (\ref{eq:einstein-equations}), and Eqs. (\ref{eq:continuity-equations}) to obtain a second-order differential equation that governs the evolution of perturbations for the cold dark matter ($w = 0$) at the linear regime 

	\begin{equation}\label{eq:ode-perturbations}
	\frac{d^2 \delta_{\rm m}}{da^2} + \left(\frac{3}{a} + \frac{1}{E}\frac{dE}{da}		\right)\delta'_{\rm m} - \frac{3}{2}\frac{\Omega_{\rm m}}{a^5E^2}\delta_{\rm m} = 0,
	\end{equation}
	where  $E(a) = H(a)/H_0$, $\Omega_{\rm m}$ is the density parameter of the dark matter today. Equation (\ref{eq:ode-perturbations}) describes how an extended viscous dark energy model tends to smooth the matter perturbations. We solve numerically this differential equation from $a_{\rm i} = 10^{-3}$ till the present time $a = 1$ $(z = 0)$ with the following initial conditions: $\delta_{\rm mi} = 1.5 \times 10^{-4}$ and $\delta'_{\rm mi} = \delta_{\rm mi}/a_{\rm i}$ which guarantees that matter perturbations are in the linear regime ($\delta_{\rm{m}0} \ll 1$).

	One can then define the linear growth function $D_{\rm m}(a)$ by $\delta_{\rm m}(a) = D_{\rm m}(a)\delta_{\rm m}(a = 1)$ which shows how much the perturbations have grown since initial moment $a_{\rm i}$. This relation is normalized in $D_{\rm m}(a = 1) = 1$. We can define the linear growth rate of the density contrast, $f$, which is related to the peculiar velocity in the linear theory, being defined by 

	\begin{equation}\label{eq:growth-rate}
	f(a) = \frac{d\ln  D_{\rm m}}{d\ln a}.
	\end{equation}
	The present galaxy surveys provide observational data for the combination $f\sigma_8$ where the linear growth rate $f$ is given by Eq. (\ref{eq:growth-rate}) and $\sigma_8$ is the root-mean-square mass fluctuation in spheres with radius $8\text{h}^{-1}\text{Mpc}$ \cite{Song2009,Huterer2015}. In the linear regime, one has \cite{Nesseris2008}

	\begin{equation}\label{eq:sigma8}
	\sigma_8(z) = \frac{\delta_{\rm m}(z)}{\delta_{\rm m}(z = 0)}\sigma_8(z = 0),
	\end{equation}
	and

	\begin{equation}\label{eq:fs8}
	f\sigma_8(z) = -(1+z)\frac{\sigma_8(z = 0)}{\delta_{\rm m}(z = 0)}\frac{d \delta_{\rm m}}{dz}.
	\end{equation}
    In the next section, we use this formalism briefly to constrain the parameters of the models studied in this work.

	In Fig. \ref{fig:delta_ms}, we show the evolution of perturbations of matter in the presence of extended viscous dark energy. We assume the same parameter combination for all models. The red line is the evolution of the $\Lambda$CDM model. For Model 1, we observe that for all combinations of $\eta$ and $q$, the values of $\delta_m$ obtained in $\log_{10} a = 0$ (today) were slightly different its for $\Lambda$CDM. Notice the evolution of Model 1 becomes different from $\Lambda$CDM in $\log_{10} a = -2.0$. Meanwhile, in Model 2, we observe that the variation of parameters renders the evolution of perturbation different from $\Lambda$CDM in $\log_{10} a = -2.5$. In late-times, the evolution is greater than that of the standard model. Finally, Model 3 has the same evolution as Model 2 in the case of a low scale factor, but today the evolution is smaller than that of $\Lambda$CDM.

	Also, we study the influence of different parameter combinations on the linear growth rate $f$. In Fig. \ref{fig:DMs} we show the evolution of the linear growth function for all models studied in this work. The line fixed in $f = 1$ means an Einstein-de Sitter Universe ($\Omega_{\rm m} = 1$). The linear growth rate for $\Lambda$CDM model tends to a constant value for high $z$ (low $a$) because $D_{\rm m} \rightarrow a$ whenever DE is very subdominant. In this case $f \rightarrow 1$. By taking Model 1, note that the abundance of DM for combination $\eta = 0.05$ and $q = 0.96$ (black line) is very similar to $\Lambda$CDM. The other combinations have an evolution higher than $\Lambda$CDM and in high redshift the growth rate converges to values greater than $f \rightarrow 1$. For Model 2, we can observe that the growth rate in late-time is slightly higher than $\Lambda$CDM. In the cases of $\eta = 0.1$, $q = 0.88$, $\eta = 0.05$ and $q = 0.85$, the evolution of $f$ are fewer than $\Lambda$CDM and the values of growth rate in high redshift is lightly different from standard model. For $\eta = 0.001$ and $q = 0.9$ (light gray), the evolution of $f$ converges faster than $\Lambda$, but to $f > 1$. In Model 3, we observe that $\eta = 0.001$ and $q = 0.98$, the growth rate is considerably lower than $\Lambda$CDM. For the others parameters combinations, the deviation from $\Lambda$CDM is fewer prominent. We can conclude that for some these models, as the $\Lambda$CDM model, the linear growth function is suppressed at low redshifts. At large redshifts, $f$ tends to a constant value since in the early periods, thus the influence of viscous dark energy can be neglected in all models.

	\section{Observational data and methodology}\label{sec:data-method}
	
	To constrain the free parameters and compare the models under consideration, we perform a Bayesian statistical analysis using recent cosmological probes, which will be summarized as follows.

	\subsection{Baryon acoustic oscillations}
	
	As the length of the Baryon acoustic oscillations (BAO) is determined through straightforward physics, it is natural to investigate the BAO scale at different times as an effective tool to constrain cosmological models. This BAO characteristic has a comoving scale of roughly $150 \rm{Mpc}$ set by the distance $r_{\rm d}$ traveled by sound waves between the end of inflation and the decoupling of baryons from photons after recombination, i.e.
	
	\begin{equation}\label{eq:scale-BAO}
		r_{\rm d}(z_{\rm d}, \textbf{p}) = \int_{z_{\rm d}}^{\infty} \frac{c_{\rm s}(z')}{H(z', \textbf{p})}dz', 
	\end{equation}
	where $z_{\rm d}$ is the redshift of the drag epoch that corresponds to the time when the baryon decouple from the photons. We assume the fitting formula of the $z_{\rm d}$  given in Ref. \cite{Eisenstein1998} to hold. $c_{\rm s}$ is the sound speed of the photon-baryon fluid given by $c_{\rm s} = \frac{c}{\sqrt{3(1 + \mathcal{R})}}$ with $\mathcal{R}= \frac{3\Omega_{\rm b}}{4\Omega_{\rm \gamma}}\frac{1}{1 + z}$.
	
	In this work, we use the BAO measurements from the last Sloan Digital Sky Survey (SDDS) collaboration release that is composed  of  data from SDSS, SDSS-II, BOSS, and eBOSS \cite{Alam:2020sor}. In this survey, the BAO feature in the line-of-sight measure the combination $D_{H}/r_{\rm d} = c/H(z, \textbf{p})r_{\rm d}$, while the transverse direction provides a measurement of $D_{\rm M}(z, \textbf{p})/r_{\rm d}$ where $D_{\rm M}$ is the comoving angular diameter distance. In a flat FLRW metric, $D_{\rm M}$ given by
	
	\begin{equation}\label{eq:comoving-distance}
		D_{\rm M}(z, \textbf{p}) = c \int_{0}^{z} \frac{dz'}{H(z', \textbf{p})}.
	\end{equation}
	The some BAO measurements are represented in spherically-averaged distance
	
	\begin{equation}\label{eq:sherical-distance}
		D_{\rm V}(z, \textbf{p}) = [zD_{\rm M}^2(z, \textbf{p})D_H(z, \textbf{p})]^{1/3}
	\end{equation}
	or more directly $D_{\rm V}(z, \textbf{p})/r_{\rm d}$.
	
	The chi-squared function $\chi^2$ related to BAO measurements is given by
	
	\begin{equation}
		\chi^2_{\rm BAO} = \sum_{i = 1}^{8} \left(\frac{X_{\rm th}(z_{\rm i}) - X_{\rm obs}(z_{\rm i})}{\sigma_{\rm i}}\right)^2,
	\end{equation} 
	where $X_{\rm th}$ is theoretical values computed by $D_{\rm V}/r_{\rm d}$, $D_{\rm M}/r_{\rm d}$ and $D_{H}/r_{\rm d}$, and  $X_{\rm obs}$ are observed values given in Table \ref{tab:BAO-data}.

	\begin{table}[h]
		\centering
		\caption{\label{tab:BAO-data} BAO distance measurements from SDDS final release.}
		\begin{tabular}{ccccc}
			\hline
			$z$ 	 & $D_{\rm V}/r_{\rm d}$  & $D_{\rm M}/r_{\rm d}$ & $D_{H}/r_{\rm d}$ & Measurement          \\ \hline 
			$0.15$   & $4.47 \pm 0.17$        & -                     & -                 & MGS                   \\
			$0.38$   & -                      & $10.23 \pm 0.17$      & $25.0 \pm 0.76$   & BOSS Galaxy           \\
			$0.51$   & -                      & $13.36 \pm 0.21$      & $22.33 \pm 0.58$  & BOSS Galaxy           \\
			$0.70$   & -                      & $17.86 \pm 0.33$      & $19.33 \pm 0.53$  & eBOSS LRG             \\
			$0.85$   & $18.33^{+0.57}_{-0.62}$ & -                     & -                 & eBOSS LRG             \\
			$1.48$   & -                      & $30.69 \pm 0.80$      & $13.26 \pm 0.55$  & eBOSS Quasar          \\
			$2.33$   & -                      & $37.6 \pm 1.9$        & $8.93 \pm 0.28$   & Ly$\alpha$-Ly$\alpha$ \\
			$2.33$   & -                      & $37.3 \pm 1.7$        & $9.08 \pm 0.34$   & Ly$\alpha$-Quasar     \\ 
			\hline
		\end{tabular}
	\end{table}
	
	\subsection{Big Bang nucleosynthesis data}
	
	We use measurement on the baryons density $\Omega_{\rm b}h^2$ based on the measurement of D/H and standard Big Bang nucleosynthesis (BBN) with modeling uncertainties \cite{Cooke2018}. The measurement is $100\Omega_{\rm b}h ^2 = 2.235 \pm 0.016$.

	\subsection{Cosmic chronometers}
	
    The cosmic chronometers (CC) is another cosmological probe obtained through the differential age method. From CC method, we can determine the Hubble parameter values at distinct redshifts based on the relative age of passively evolving galaxies (see \cite{GmezValent2018} and the references therein).
	
	The chi-squared function $\chi^2$ for CC is given by
	
	\begin{equation}
		\chi^2_{\rm CC} = \sum_{i = 1}^{31} \left(\frac{H_{\rm th}(z_{\rm i}) - H_{\rm obs}(z_{\rm i})}{\sigma_{\rm i}}\right)^2,
	\end{equation} 
	where  $H_{\rm th}(z_{\rm i})$ is computed from the Eqs. (\ref{eq:hubble1}), (\ref{eq:hubble2}) and (\ref{eq:hubble3}), and $\sigma_{\rm i}$ is the uncertainty associated with $H_{\rm obs}(z_{\rm i})$ measurement. In Table \ref{tab:CC-data}, we use the $31$ observational points of the Hubble parameter in the redshift range $0.07 < z < 1.96$ related in Refs. \cite{Farooq2017,GmezValent2018}.
	
	\begin{table}[h]
		\centering
		\caption{\label{tab:CC-data} Thirty-one observational values of $H(z)$ in [km/s/Mpc] and associated uncertainties obtained using the differential-age technique.}
			\begin{tabular}{cc|cc}
				\hline
				$z$ &  $H(z) \pm \sigma_{H}$ & $z$ &  $H(z)\pm \sigma_{H}$
				 \\ \hline
				$0.07$ & $69.0\pm 19.6$ & $0.4783$ & $80.9\pm 9.0$
				\\ 
				$0.09$ & $69.0\pm 12.0$ & $0.48$ & $97.0\pm 62.0$ 
				\\
				$0.12$ & $68.6\pm 26.2$ & $0.5929$ & $104.0\pm 13.0$
				\\
				$0.17$ & $83.0\pm 8.0$  & $0.6797$ & $92.0\pm 8.0$
				\\
				$0.1791$ & $75.0\pm 4.0$ & $0.7812$ & $105.0\pm 12.0$ 
				\\
				$0.1993$ & $75.0\pm 5.0$ & $0.8754$ & $125.0\pm 17.0$
				\\
				$0.2$ & $72.9\pm 29.6$ & $0.88$ & $90.0\pm 40.0$
				\\
				$0.27$ & $77.0\pm 14.0$ & $0.9$ & $117.0\pm 23.0$
				\\
				$0.28$ & $88.8\pm 36.6$ & $1.037$ & $154.0\pm 20.0$
				\\
				$0.3519$ & $83.0\pm 14.0$ & $1.3$ & $168.0\pm 17.0$ 
				\\
				$0.3802$ & $83.0\pm 13.5$ & $1.363$ & $160.0\pm 33.6$ 
				\\
				$0.4$ & $95.0\pm 17.0$ & $1.43$ & $177.0\pm 18.0$ 
				\\
				$0.4004$ & $77.0\pm 10.2$ & $1.53$ & $140.0\pm 14.0$
				\\
				$0.4247$ & $87.1\pm 11.2$ & $1.75$ & $202.0\pm 40.0$  
				\\
				$0.4497$ & $92.8\pm 12.9$ & $1.965$ & $186.5\pm 50.4$
				\\
				$0.47$ & $89.0\pm 49.6$ & &
			\\ \hline	
			\end{tabular}
	\end{table}

	\subsection{Cosmic microwave background distance priors}
	
	The cosmic microwave background (CMB) is an important observable in cosmology due to knowledge about the early Universe as well as its capacity to constrain cosmological parameters. In Refs. \cite{Kosowsky2002, Elgary2007, Wang2007,Efstathiou1999,Mukherjee2008,Chen2019} show the possibility to compress CMB likelihood in few numbers: CMB shift parameter $\mathcal{R}$, the angular scale of the sound horizon at last scattering $\ell_A$, and baryon density today $\Omega_{b}h^2$.
	
	In a spatially flat Universe, we define the shift parameter $\mathcal{R}$ as
	
	\begin{equation}\label{R}
		\mathcal{R}(\textbf{p}) = \sqrt{\Omega_{\rm m}H^2_0}\frac{D_{\rm M}(z_\star, \textbf{p})}{c},
	\end{equation}
	and the angular scale of the sound horizon at last scattering is given by
	
	\begin{equation}\label{la}
		\ell_A(\textbf{p}) = \pi \frac{D_{\rm M}(z_\star, \textbf{p})}{r_s(z_\star, \textbf{p})},
	\end{equation}
	where $r_s(z_\star, \textbf{p})$ is comoving size of the sound horizon, Eq. (\ref{eq:scale-BAO}), computed at redshift at the photon decoupling epoch $z_\star$ \cite{Hu1996}. Then $\chi^2$ function of the CMB priors is defined as
	
	\begin{equation}
		\chi^{2}_{\rm CMB} = \textbf{X}^{T}_{\rm CMB}\cdot \textbf{C}^{-1}_{\rm CMB} \cdot\textbf{X}_{\rm CMB},
	\end{equation}
	where $ \textbf{X}_{\rm CMB} = (\mathcal{R}^{\rm th}, \ell_A^{\rm th}, \Omega_{\rm b} h^2) - (\mathcal{R}^{\rm obs}, \ell_A^{\rm obs}, \Omega_{\rm b} h^{2})$ and $\textbf{C}$ is covariance matrix. We use CMB distance priors from Planck 2018 release \cite{Chen2019}.

	\subsection{Local value of the \texorpdfstring{$H_0$}{}}
	
	We consider the measurement of the local value of the Hubble constant by the Hubble Space Telescope (HST): $H_0 = 74.03 \pm 1.42 $ km/s/Mpc \cite{Riess2019}.
	
	\subsection{\texorpdfstring{$f\sigma_8$}{} measurements}
	
	Most growth rate measurements are achieved using peculiar velocities acquire from redshift-space distortions (RSD) measurements described in galaxy redshift surveys \cite{Kaiser1987}. In general, such surveys can supply measurements of the perturbations in terms of galaxy density $\delta_{\rm g}$, which are related to matter perturbations through the bias parameter $b$ as $\delta_{\rm g} = b \delta_{\rm m}$ \cite{Nesseris2017}. A more trustworthy combination is the product $f\sigma_8$ as it is independent of the bias and may obtain using either weak leasing or RSD. The measured value of $f\sigma_8$ depends on the fiducial cosmology, so to do the measurement model-independent, one needs to re-scale by a factor \cite{Nesseris2017}
	
	\begin{equation}\label{eq:ratio}
		ratio(z, \textbf{p}) = \frac{H(z, \textbf{p})D_{\rm A}(z, \textbf{p})}{H^{\rm fid}(z)D_{A}^{\rm fid}(z)},
	\end{equation}
	where $D_{A}^{\rm fid}$ is the angular distance of the fiducial model, and the values of the fiducial cosmology can found in data point references. Then, invoking the correction gives by Eq. (\ref{eq:ratio}) and multiplying it on the theoretical prediction of $f\sigma_8$, we can be sure which our data points are model-independent.
	
	We use $22$ $f\sigma_8$ measurements from RSD observations according to the Gold-$2018$ compilation \cite{Sagredo2018,Arjona:2020yum}. In Table \ref{tab:RSD-data} are shown the RSD measurements given in different redshifts. The values of matter density for fiducial $\Lambda$CDM model are given too. Some $f\sigma_8$ points are correlated with each other (points $13-15$ and $19-22$). These data points are the three from WiggleZ \cite{Blake:2012pj}, and the four points from SDDS \cite{Zhao:2018jxv}. The related covariance matrix to WiggleZ data is given by
	
	\begin{equation}\label{WiggleZCov}
		\textbf{C}_{\rm WiggleZ}= 10^{-3} \times
		\left(
		\begin{array}{ccc}
			6.400 & 2.570 & 0.000 \\
			2.570 & 3.969 & 2.540 \\
			0.000 & 2.540 & 5.184 \\
		\end{array}
		\right),
	\end{equation}
	and the covariance matrix of the SDSS data is
	\begin{equation}\label{SDSS4Cov}
		\textbf{C}_{\rm SDSS-IV}= 10^{-2} \times
		\left(
		\begin{array}{cccc}
			3.098 & 0.892 &  0.329 & -0.021\\
			0.892 & 0.980 & 0.436 & 0.076\\
			0.329 & 0.436 &  0.490   & 0.350 \\
			-0.021 & 0.076 & 0.350 & 1.124
		\end{array}
		\right).
	\end{equation}

	\begin{table}[h]
		\centering
		\caption{\label{tab:RSD-data} Compilation of the $f\sigma_8(z)$ measurements used in this work and the reference matter density parameter $\Omega_{\rm m}$ and related references.}
			\begin{tabular}{ccccc}
				\hline
				$z$     & $f\sigma_8(z)$ & $\sigma_{f\sigma_8}$  & $\Omega_{\rm m}$ & Ref. \\ \hline
				0.02    & 0.428 & 0.0465  & 0.3 & \cite{Huterer:2016uyq}   \\
				0.02    & 0.398 & 0.065   & 0.3 & \cite{Turnbull:2011ty},\cite{Hudson:2012gt} \\
				0.02    & 0.314 & 0.048   & 0.266 & \cite{Davis:2010sw},\cite{Hudson:2012gt}  \\
				0.10    & 0.370 & 0.130   & 0.3 & \cite{Feix:2015dla}  \\
				0.15    & 0.490 & 0.145   & 0.31 & \cite{Howlett:2014opa}  \\
				0.17    & 0.510 & 0.060   & 0.3 & \cite{Song:2008qt}  \\
				0.18    & 0.360 & 0.090   & 0.27 & \cite{Blake:2013nif} \\
				0.38    & 0.440 & 0.060   & 0.27 & \cite{Blake:2013nif} \\
				0.25    & 0.3512 & 0.0583 & 0.25 & \cite{Samushia:2011cs} \\
				0.37    & 0.4602 & 0.0378 & 0.25 & \cite{Samushia:2011cs} \\
				0.32    & 0.384 & 0.095  & 0.274 & \cite{Sanchez:2013tga}   \\
				0.59    & 0.488  & 0.060 & 0.307115 & \cite{Chuang:2013wga} \\
				0.44    & 0.413  & 0.080 & 0.27 & \cite{Blake:2012pj} \\
				0.60    & 0.390  & 0.063 & 0.27 & \cite{Blake:2012pj} \\
				0.73    & 0.437  & 0.072 & 0.27 & \cite{Blake:2012pj} \\
				0.60    & 0.550  & 0.120 & 0.3 & \cite{Pezzotta:2016gbo} \\
				0.86    & 0.400  & 0.110 & 0.3 & \cite{Pezzotta:2016gbo} \\
				1.40    & 0.482  & 0.116 & 0.27 & \cite{Okumura:2015lvp} \\
				0.978   & 0.379  & 0.176 & 0.31 & \cite{Zhao:2018jxv} \\
				1.23    & 0.385  & 0.099 & 0.31 & \cite{Zhao:2018jxv} \\
				1.526   & 0.342  & 0.070 & 0.31 & \cite{Zhao:2018jxv} \\
				1.944   & 0.364  & 0.106 & 0.31 & \cite{Zhao:2018jxv} \\
				\hline

			\end{tabular}

	\end{table}

    The chi-squared function $\chi^2$ for RSD is defined as
	
	\begin{equation}
	\chi^2_{\rm RSD}=  \textbf{X}_{\rm RSD}^T \cdot \textbf{C}^{-1} \cdot \textbf{X}_{\rm RSD},
	\end{equation}
	where $\textbf{C}$ is covariance matrix and $\textbf{X}$ is given by
	
	\begin{equation}
		\textbf{X} = f\sigma_8^{\rm obs,i} -  ratio(z_i)f\sigma_8^{\rm th,i}
	\end{equation}
	where $f\sigma_8^{\rm obs,i}$ is the observed data, $ratio(z_i)$ is given by Eq. (\ref{eq:ratio}) and $f\sigma_8^{\rm th,i}$ is theoretical prediction computed by Eq. (\ref{eq:fs8}).

	\subsection{Type Ia Supernovae}
	
    The type Ia supernovae (SNe Ia) data are the most relevant measurements in cosmology and comprise important evidence of the cosmic acceleration. Besides that, they are a powerful probe to constraint cosmological parameters. The Pantheon compilation consists of $1048$ SNe Ia objects in the redshift interval $0.01 < z < 2.3$. According to Ref. \cite{Scolnic2018}, we use this sample with stretch and color luminosity parameters null. 
	
	The theoretical distance modulus $\mu_{\rm th}$ for a given SNe Ia in redshift is defined by
	
	\begin{equation}
		\mu_{\rm th}(z, \textbf{p}) = 5\log_{10}\frac{d_{\rm L}(z, \textbf{p})}{\rm Mpc} + 25,
	\end{equation}
	where $d_{\rm L} = (c/H_{0})D_{\rm L}$ is the luminosity distance. The Hubble-free luminosity distance is given by
	
	\begin{equation}
		D_{\rm L}(z, \textbf{p}) = (1 + z_{\rm hel})\int_{0}^{z_{\rm CMB}}\frac{dz'}{E(z', \textbf{p})},
	\end{equation} 
	where $E(z) = H(z)/H_{0}$, $z_{\rm hel}$ and $z_{\rm CMB}$ are the dimensionless Hubble parameter, heliocentric and CMB frame redshifts, respectively.
	
	The observed distance modulus reads
	
	\begin{equation}
		\mu_{\rm obs} = m_{\rm B} - M_{\rm B},
	\end{equation}
	where $m_{\rm B}$ is the observed peak magnitude in the rest frame of the B band and $M_{\rm B}$ is the absolute magnitude nuisance of the SNe Ia. The $\chi^2$ function is
	
	\begin{equation}
		\chi^{2}_{\rm SNe Ia} = \textbf{m}^{T}\textbf{C}^{-1}\textbf{m},
	\end{equation}
	where $C$ is the covariance matrix composed by $\textbf{C} = \textbf{D}_{\rm stat} + \textbf{C}_{\rm sys}$ where $\textbf{D}_{\rm stat}$ is the diagonal covariance matrix of statistical uncertainties, and $\textbf{C}_{\rm sys}$ is the matrix of systematic errors and $\textbf{m} = m_{\rm b}(z_{\rm i}) - m_{\rm th}(z_{\rm i}, \textbf{p})$, where
	
	\begin{equation}
		m_{\rm th}(z, \textbf{p})  = 5\log_{10}D_{\rm L}(z, \textbf{p}) + M,
	\end{equation}
	in which $H_0$ and $M_{\rm B}$ can be absorbed into $M$. The nuisance parameter $M$ could be marginalized following Ref. \cite{Conley2010}. After the marginalization over $M$, we define the quantities
	
	\begin{eqnarray}
		a_1 & = &  \Delta\textbf{m}^{T}\cdot \textbf{C}^{-1}\cdot \Delta\textbf{m},\\
		a_2 & = &  \Delta\textbf{m}^{T}\cdot \textbf{C}^{-1}\cdot \textbf{1}, \\
		a_3 & = &  \textbf{1}^{T}\cdot \textbf{C}^{-1}\cdot \textbf{1},
	\end{eqnarray}
	where $ \Delta\textbf{m} = m_{\rm B} - m_{\rm th}$ and $\textbf{1}$ is identity matrix; finally, the $\chi^2$ function reads
	
	\begin{equation}
		\chi^{2}_{\rm SNe Ia} = a_1 - \frac{a_2^2}{a_3} + \ln \frac{a_3}{2\pi}.
	\end{equation}
	
	For joint analysis, we assume the sum of all $\chi^2$ functions: $\chi^2_{\rm joint} = \chi^2_{\rm BAO} + \chi^2_{\rm BBN} + \chi^2_{\rm H} + \chi^2_{\rm CMB} + \chi^2_{\rm H_{0}} + \chi^2_{\rm f\sigma_8} + \chi^2_{\rm SNeIa}$.
	
	In our analysis are associated with four different steps. First, we consider the combination of CMB priors + SNe Ia + BAO + CC in order to constrain the set of parameters $\textbf{p} = (H_{0}, \Omega_{\rm b}, \Omega_{\rm c}, \Omega_{\rm m}, \eta, q)$. After the first statistical analysis, we add the $H_0$ local measurement to investigate the effect of the local data in our models. Third step, we use all background data jointed to the growth rate data to constrain the parameters $\textbf{p} = (H_{0}, \Omega_{\rm b}, \Omega_{\rm c}, \Omega_{\rm m}, \eta, q, \sigma_8)$. And the last step, we take into account all background measurements plus $H_0$, and the growth rate data.

	To implement the statistical analysis, we took the public package \textsf{MultiNest} \cite{Feroz2008,Feroz2019,Buchner2014} through the \textsf{PyMultiNest} interface \cite{Buchner2014}. To perform this analysis, we choose uniform priors about the free parameters of the models investigated. These priors are $55.84 \leq H_0 \leq 90.64$, $0.001 \leq \Omega_{\rm b}h^2 \leq 0.1$, $0.005 \leq, \Omega_{\rm c}h^2 \leq 0.99$, $ 0.001\leq \Omega_{\rm m} \leq 0.99$, $0 \leq \eta \leq 1 $, $0.8 \leq q \leq 1.4$, and $0.4 \leq\sigma_8 \leq 1.2$. The previous distribution on $ \eta$ in this work is different from the previous one in which we assume a Gaussian prior centered in zero \cite{Silva2019}. We fix the radiation density parameter in $\Omega_{\text{r}} h^2 = 1.698 \Omega_\gamma$ with $\Omega_\gamma = 2.469 \times 10^{-5}h^2$ where $H_0 = 100 h \ \text{km} \ \text{s}^{-1} \text{Mpc}^{-1}$.

	\section{Results and discussion}\label{sec:results}

    \begin{table*}[t]
		\caption{\label{tab:constraints1} Constraints on the parameters for $\Lambda$CDM, q-$\Lambda$CDM Models 1, 2 and 3 considering the combination CMB priors + SNe Ia + BAO + BBN + CC.}
		\begin{tabular}{lccccc}
			\hline
			& Model 1 & Model 2 & Model 3  & q-$\Lambda$CDM & $\Lambda$CDM\\ \hline
			$H_0$                  & $69.3\pm 1.7 $                 & $69.3\pm 1.7 $              & $69.2\pm 1.7$ & $69.1\pm 1.7 $ &  $67.38\pm 0.80            $      \\
			$\Omega_{\rm b}h^2$    & $0.02235\pm 0.00016$			& $0.02235\pm 0.00016$        & $0.02235\pm 0.00016$ & $0.02235\pm 0.00016$ & $0.02236\pm 0.00016 $ \\
			$\Omega_{\rm c}h^2$	   & $0.1247 \pm 0.0128 $           & $0.1189 \pm 0.0209$         & $0.1215 \pm 0.0117$ & $0.1114 \pm 0.0081$ & $0.1173\pm 0.0069$ \\ 
			$\Omega_m$             & $0.307 \pm 0.066$              & $0.295 \pm 0.022$           & $0.301 \pm 0.031 $  & $0.281 \pm 0.025$ & $0.307\pm 0.012            $ \\
			$\eta$                 & $ < 0.0505     $               & $ < 0.0134  $                 & $ < 0.104  $              & -  & -  \\
			$q$                    & $0.968\pm 0.044$               & $0.969\pm 0.045  $           & $0.977\pm 0.046 $   & $0.951  \pm 0.043 $    & -     \\ \hline
		\end{tabular}
	\end{table*}

	\begin{table*}[t]
		\caption{\label{tab:constraints2} Constraints on the parameters for $\Lambda$CDM, q-$\Lambda$CDM Models 1, 2 and 3 considering the combination CMB priors + SNe Ia + BAO + BBN + CC + RSD.}
		\begin{tabular}{lccccc}
			\hline
			& Model 1 & Model 2 & Model 3  & q-$\Lambda$CDM & $\Lambda$CDM\\ \hline
			$H_0$                  & $69.3\pm 1.7 $                 & $69.7\pm 1.7  $           & $69.6\pm 1.7 $       & $69.4\pm 1.7  $ &  $67.35\pm 0.79            $      \\
			$\Omega_{\rm b}h^2$    & $0.02235\pm 0.00016$  			& $0.02235\pm 0.00016$      & $0.02234\pm 0.00016$ & $0.02235\pm 0.00016$ & $0.02237\pm 0.00016 $ \\
			$\Omega_{\rm c}h^2$	   & $0.1199 \pm 0.0114 $           & $0.1155 \pm 0.0091$       & $0.1169 \pm 0.0097$  & $0.1093 \pm 0.008$& $0.1157\pm 0.00669$ \\ 
			$\Omega_m$             & $0.293 \pm 0.029$              & $0.285\pm 0.025$          & $0.288\pm 0.026 $    & $0.274  \pm 0.023$ & $0.304\pm 0.012            $ \\
			$\eta$                 & $ < 0.0431     $               & $ < 0.0120 $              & $ < 0.0861    $      & -  & -  \\
			$q$                    & $0.953\pm 0.043$               & $0.956\pm 0.043$          & $0.960\pm 0.043 $    & $0.941\pm 0.042$    & -     \\ 
			$\sigma_8$			   & $0.774\pm 0.047$				& $0.785\pm 0.043$			& $0.780\pm 0.044 $	   & $0.803\pm 0.041$  & $0.764\pm 0.030$ \\	\hline
		\end{tabular}
	i\end{table*}

	This section presents the main results from the statistical analysis obtained through observational data and assumes the scenarios defined in section \ref{sec:phenomenology}. In the first subsection, we will show the parameter constraints achieved by statistical analysis according to steps described in section \ref{sec:data-method}. In second subsection, we will implement the model comparison between the models studied and, in last subsection, we will study cosmic age crisis.

	\subsection{Parameter Constraints}

	\begin{table*}[t]
		\caption{\label{tab:constraints3} Constraints on the parameters for $\Lambda$CDM, q-$\Lambda$CDM Models 1, 2 and 3 considering the combination CMB priors + SNe Ia + BAO + BBN + CC + $H_0$.}
		\begin{tabular}{lccccc}
			\hline
			& Model 1 & Model 2 & Model 3  & q-$\Lambda$CDM & $\Lambda$CDM\\ \hline
			$H_0$                  & $72.1\pm 1.1 $               & $72.1\pm 1.1  $        & $72.1\pm 1.1 $           & $72.1   \pm 1.1 $ &  $69.09\pm 0.73         $      \\
			$\Omega_{\rm b}h^2$    & $0.02235\pm 0.00016$	      & $0.02235\pm 0.00016 $  & $0.02235\pm 0.00016$     & $0.02235\pm 0.00016$& $0.02240\pm 0.00016        $  \\
			$\Omega_{\rm c}h^2$	   & $0.1224 \pm 0.0139 $         & $0.1154 \pm 0.0104$    & $0.1182 \pm 0.0116 $     & $0.1068 \pm 0.0073$& $0.1260\pm 0.0069          $\\ 
			$\Omega_m$             & $0.278 \pm 0.028$            & $0.265\pm 0.023$       & $0.271 \pm 0.02$         & $0.249  \pm 0.0175$ & $0.311\pm 0.012           $ \\
			$\eta$                 & $ < 0.0604      $            & $ < 0.0155 $           & $ < 0.115    $           & -  & -  \\
			$q$                    & $0.911\pm 0.038$             & $0.911\pm 0.038   $    & $0.919\pm 0.042 $        & $0.885\pm 0.033 $    & -     \\ \hline
		\end{tabular}
	\end{table*}

	\begin{table*}[t]
	\caption{\label{tab:constraints4} Constraints on the parameters for $\Lambda$CDM, q-$\Lambda$CDM Models 1, 2 and 3 considering the combination CMB priors + SNe Ia + BAO + BBN + CC + $H_0$ + RSD.}
	\begin{tabular}{lccccc}
		\hline
		& Model 1 & Model 2 & Model 3  & q-$\Lambda$CDM & $\Lambda$CDM\\ \hline
		$H_0$                  & $72.2\pm 1.1 $                 & $72.2\pm 1.1  $           & $72.2\pm 1.1$       & $72.1\pm 1.1 $ &  $69.06\pm 0.73          $      \\
		$\Omega_{\rm b}h^2$    & $0.02235\pm 0.00016$  			& $0.02235\pm 0.00016$      & $0.02234\pm 0.00016$ & $0.02235\pm 0.00016        $ & $0.02240\pm 0.00016 $ \\
		$\Omega_{\rm c}h^2$	   & $0.1183 \pm 0.0114 $           & $0.1136 \pm 0.0091$       & $0.1156 \pm 0.0103 $  &$0.1059\pm 0.0071  $ & $0.1243\pm 0.0068$ \\ 
		$\Omega_m$             & $0.270 \pm 0.024$              & $0.261 \pm 0.020$         & $0.265 \pm 0.022 $    & $0.247 \pm 0.017$         & $0.308\pm 0.012             $ \\
		$\eta$                 & $ < 0.0506     $               & $ < 0.0141 $              & $ < 0.102     $      & -  & -  \\
		$q$                    & $0.902\pm 0.035$               & $0.906\pm 0.037$          & $0.911\pm 0.039  $    & $0.882\pm 0.032 $    & -     \\ 
		$\sigma_8$			   & $0.803\pm 0.047 $				& $0.814\pm 0.042 $			& $0.809\pm 0.045 $	   & $0.841\pm 0.039            $ & $0.763\pm 0.0290$ \\	\hline
	\end{tabular}
	\end{table*}

\begin{figure*}
		\centering
		\includegraphics[scale=0.20]{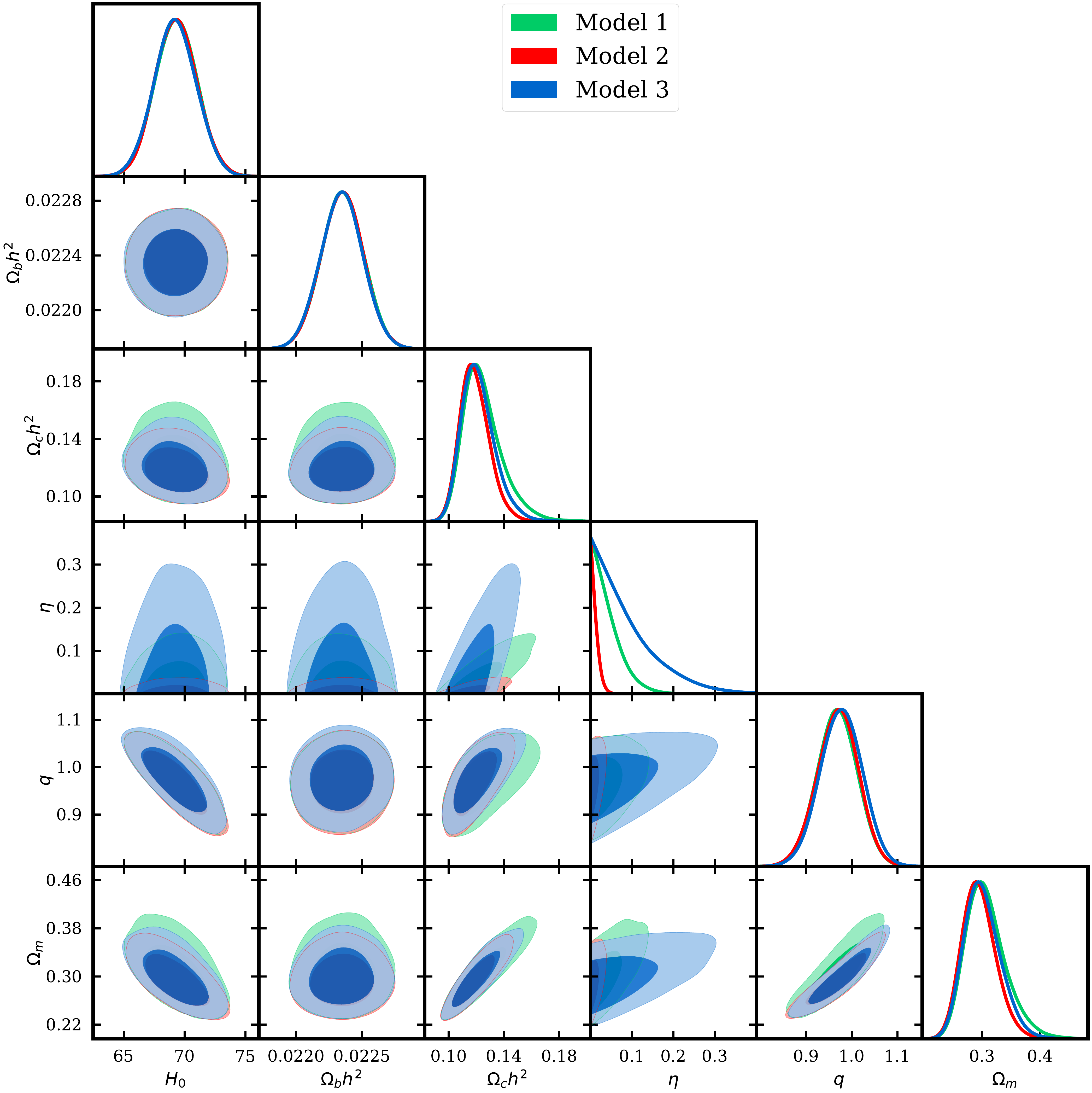}
		\caption{\label{fig:contours1} 2-dimensional confidence contours and 1-dimension posterior distributions on the free parameters of each model analyzed using the cosmological probes. The panel shows the results for all models considering  CMB priors + SNe Ia + BAO + BBN + CC.}
	\end{figure*}

	The results obtained for extended viscous dark energy models are summarized in Table \ref{tab:constraints1} when we show the means and $1\sigma$ uncertainties of the free parameters considering the combination CMB priors + SNe Ia + BAO + BBN + CC. At the same time, to make a comparison of the extended viscous dark energy models with the $\Lambda$CDM, in Table \ref{tab:constraints1}, we show the constraints on the  $\Lambda$CDM and q-$\Lambda$CDM. Additionally, in  Fig. \ref{fig:contours1}, we present the corresponding contour plots for each model. Our statistical analyses show that for all viscous models, the Hubble constant $H_0$ takes a value slightly greater than one obtained by the $\Lambda$CDM model. The baryon density values achieved by all models were unaffected. The cold dark matter density values obtained were compatible with $\Lambda$CDM in $1 \sigma$ CL. For the three extended viscous dark energy models, we obtain the following values for bulk viscosity parameter $\eta < 0.0505$, $0.0134$ and $0.104$, respectively. These values are the $2\sigma$ upper bounds and suffer the influence of the prior distribution in which we consider positive values for $\eta$. The parameter $q$ values show that cosmological data and the models are slightly compatible ($2\sigma$ CL) with the extended framework. Then, from the combination CMB priors + SNe Ia + BAO + BBN + CC, the $\Lambda$CDM is recovery with $1\sigma$ confidence. By regarding the $H_0$ tension, notice our results decreased the tension in $2 \sim 3 \sigma$ with local measurement \cite{Riess2019}.

\begin{figure*}
		\centering
		\includegraphics[scale=0.20]{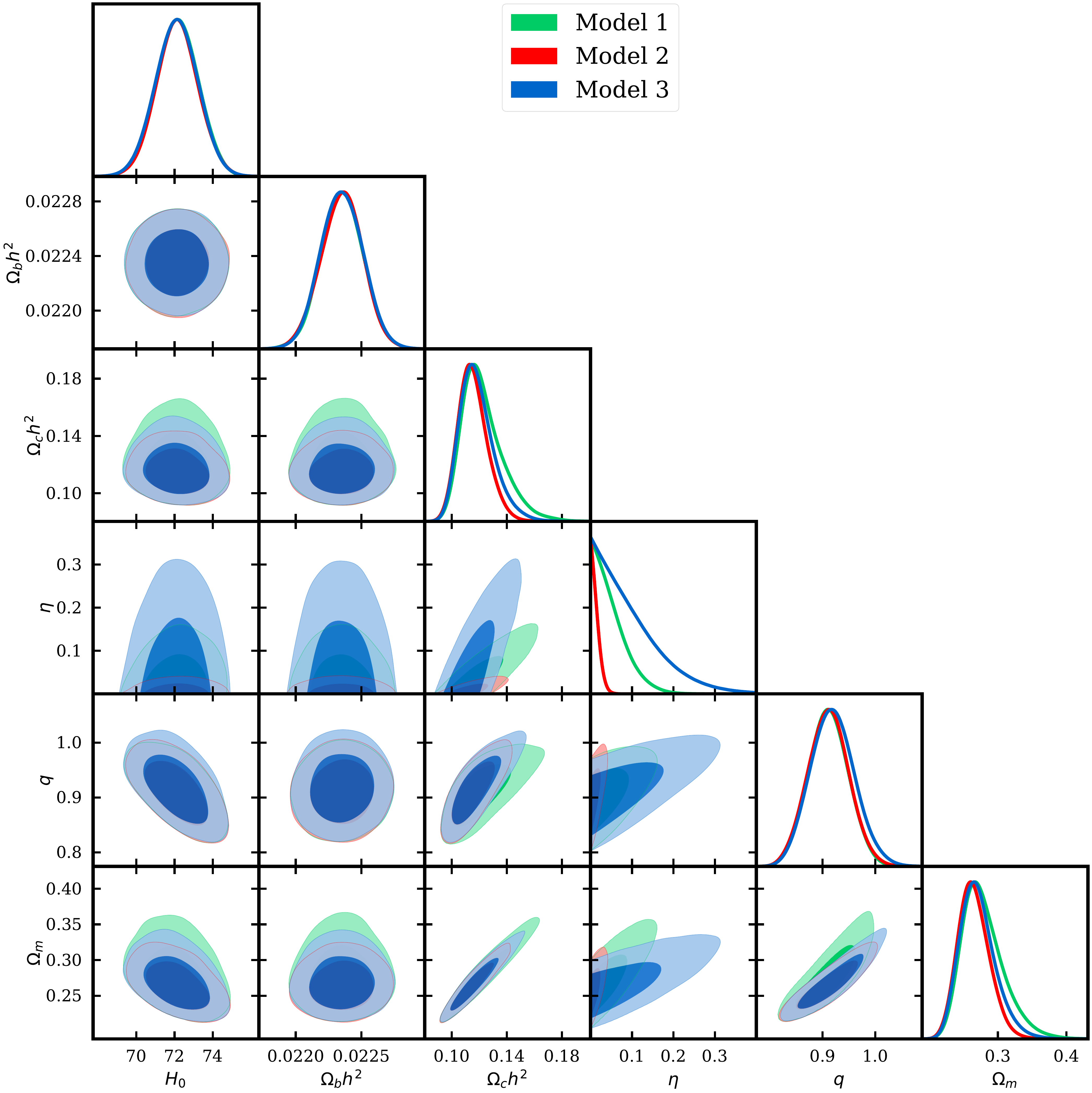}
		\caption{\label{fig:contours3} 2-dimensional confidence contours and 1-dimension posterior distributions on the free parameters of each model analyzed using the cosmological probes. The panel shows the results for all models considering  CMB priors + SNe Ia + BAO + BBN + CC + RSD.}
	\end{figure*}

	Now by addition of the RSD data, we achieved the results shown in Table \ref{tab:constraints2} and Fig. \ref{fig:contours3}. The statistical analyses show the value of Hubble constant $ H_0 $ is little different ones previously obtained to all models. The values of the matter density are consistent in $ 1\sigma$ confidence with those obtained previously. Notice the values obtained for bulk viscosity parameter $\eta$ decreased approximately $\approx 14 \%$ (Model 1), $\approx 10 \%$ (Model 2) and $\approx 17 \%$ (Model 3).  With the addition of the RSD data further restricted the bulk viscosity. The same behavior occurs in relation to nonadditive parameter $q$, $\approx 2\%$ for all models. Let us focus on $\sigma_8$, which plays a particularly relevant role in formation structure.  From Table \ref{tab:constraints2}, we can read off $\sigma_8 = 0.764 \pm 0.030$ for  the  $\Lambda$CDM  model, whereas the Model 1 prediction is $\sigma_8 = 0.774 \pm 0.047$, Model 2 is $\sigma_8 = 0.785 \pm 0.043$, and Model 3 is $\sigma_8 = 0.780 \pm 0.044$. The values obtained for all viscous models were compatible for $1\sigma$ CL with $\Lambda$CDM. Notice that our results for $\sigma_8$ indicate that the tension of $\sigma_8$ is reduced to $0.780\sigma$ $0.598\sigma$ and $0.698\sigma$, for Models 1, 2 and 3, respectively.

\begin{figure*}
		\centering
		\includegraphics[scale=0.20]{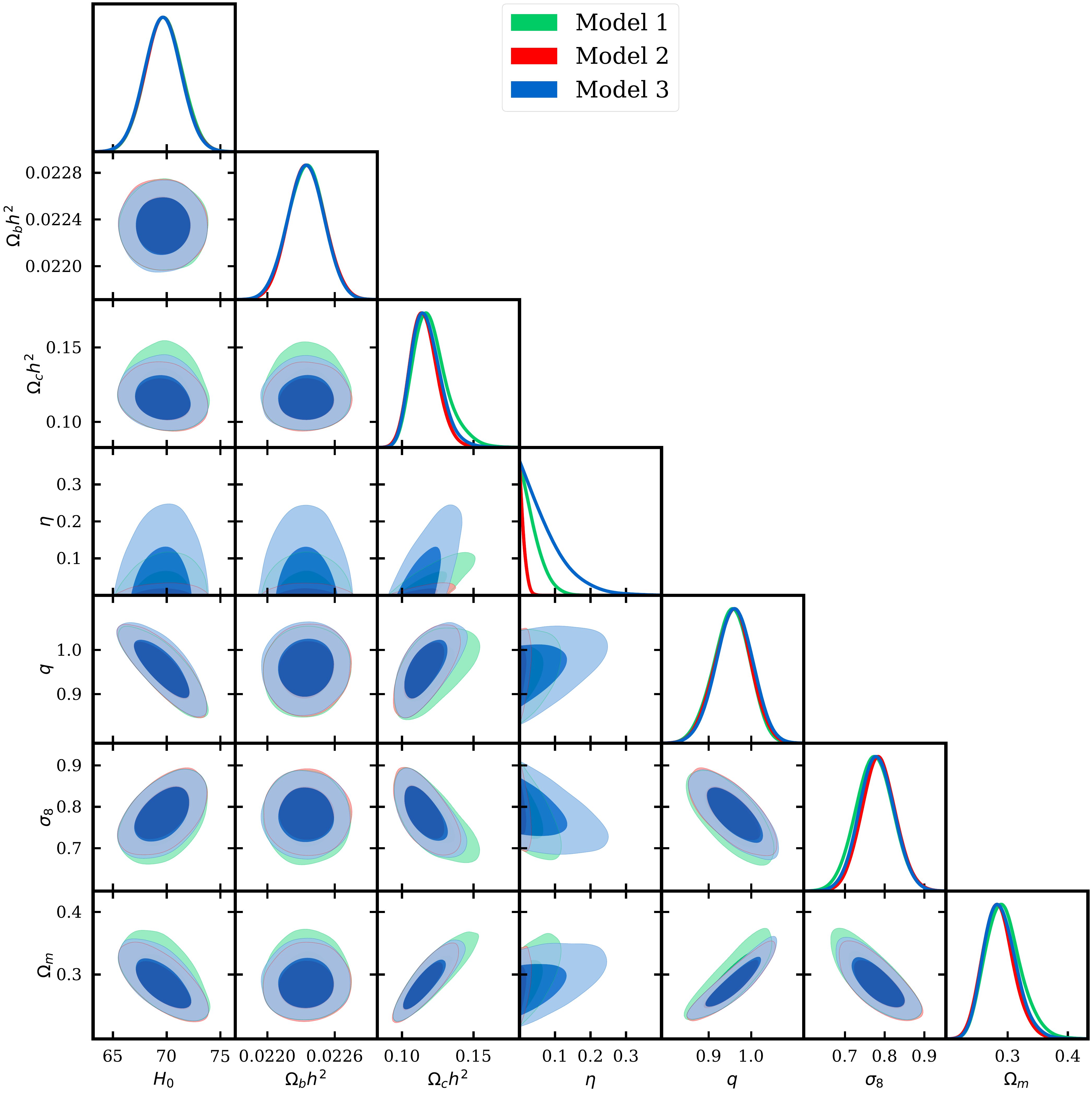}
		\caption{\label{fig:contours2} 2-dimensional confidence contours and 1-dimension posterior distributions on the free parameters of each model analyzed using the cosmological probes. The panel shows the results for all models considering CMB priors + SNe Ia + BAO + BBN + CC + $H_0$.}
	\end{figure*}

	In the following, we assume the inclusion of the $H_{0}$ local, Table \ref{tab:constraints3}, and Fig. \ref{fig:contours2}. Notice that the constraints of the parameters $H_0$, and $\Omega_{\rm c}h^2$ have deviated to higher and lower fit values, respectively, when compared to the previous analysis. However, the current analysis and the previous one are compatible with $1\sigma$ CL. By considering the bulk viscosity parameter, we obtain $\eta < 0.0604$, $0.0155$, $0.115$. With the addition of the $H_{0}$ local, the values of $\eta$ increased $\approx 19  \%$ and $15 \%$ for Models 1 and 2, respectively, and $\approx 10 \%$ for Model 3. Finally, the value of the nonadditive parameter $q$ decreased $\approx 4 \%$. Notice the values obtained are compatible at $1\sigma$ CL with previous analysis. Now, about the tension of $H_0$, the extended viscous dark energy models relieve this tension in  $\approx 1 \sigma$ with local measurement \cite{Riess2019}.

\begin{figure*}
		\centering
		\includegraphics[scale=0.20]{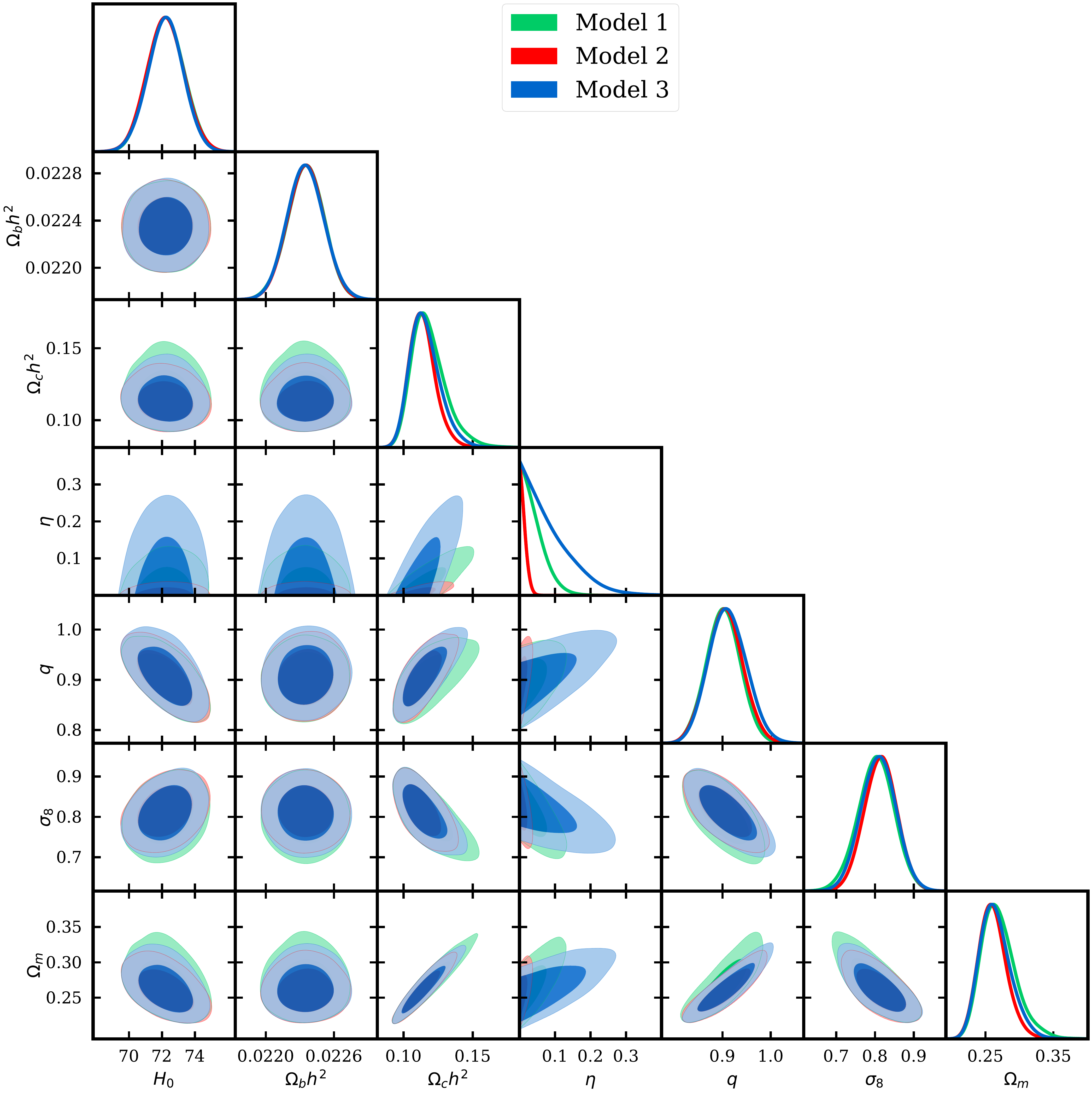}
		\caption{\label{fig:contours4} 2-dimensional confidence contours and 1-dimension posterior distributions on the free parameters of each model analyzed using the cosmological probes. The panel shows the results for all models considering CMB priors + SNe Ia + BAO + BBN + CC + RSD + $H_0$.}
	\end{figure*}

	In the last analysis, we included the RSD data; see Table \ref{tab:constraints4}, and Fig. \ref{fig:contours4}. The results for $H_0$ were not influenced by adding RSD data. The matter density is slightly different for all models but, it agrees at $ 1\sigma$ CL with previous analysis. The values of the bulk viscosity and the nonadditive parameter decreased if compared to the last data combination. Finally, let us concentrate on $\sigma_8$ again. With the addition of $H_{0}$ local, the values obtained for $\sigma_8$ are higher than the previous ones and, in this way, further relieve the tension of $\sigma_8$.
	
	\subsection{Model Selection}
	
	\begin{figure}[H]
		\centering
		\includegraphics[width=8.0cm]{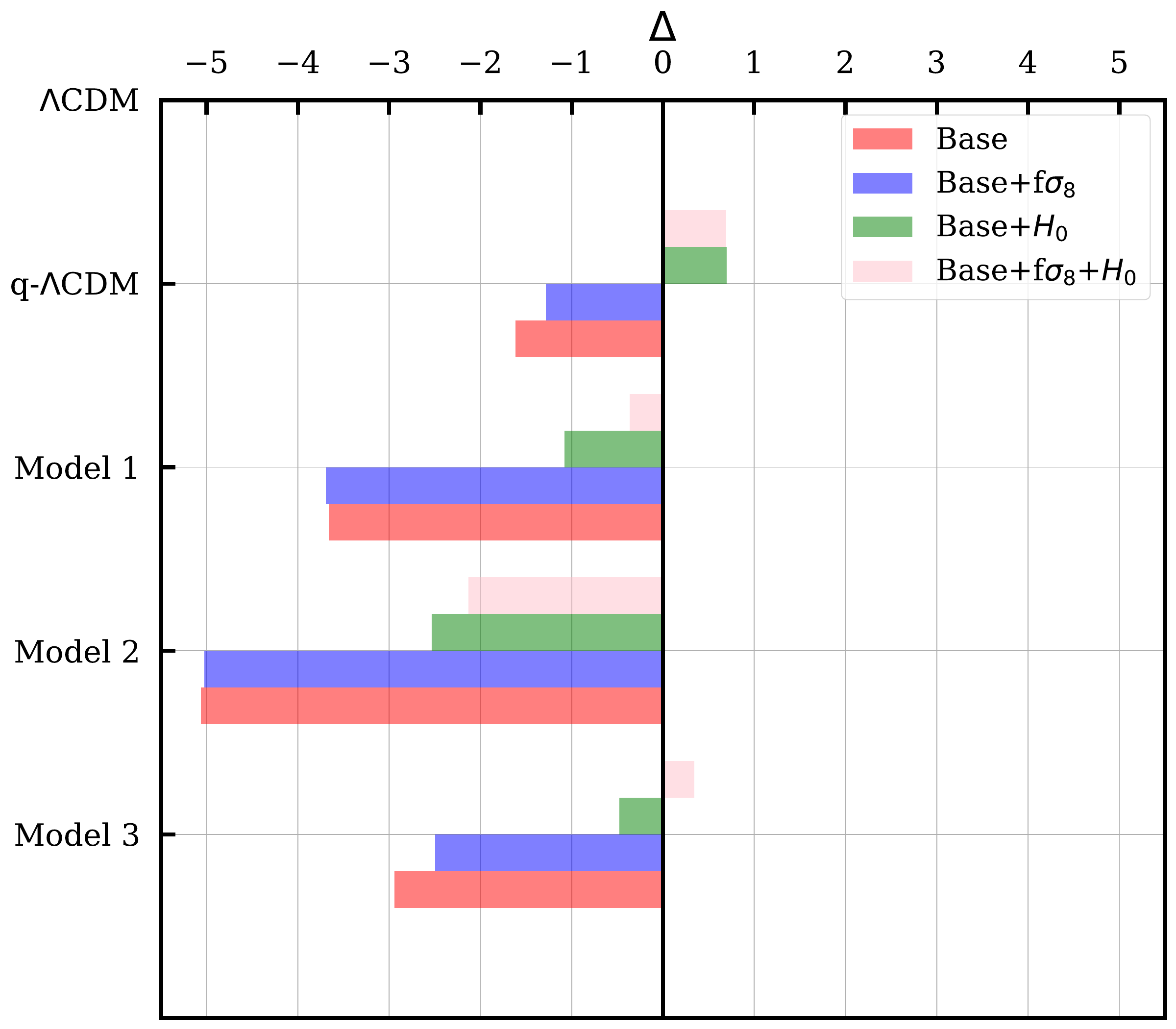}
		\caption{\label{fig:bayes-factor} Figure shows the Bayes factor between each extended viscous model and $\Lambda$CDM considering the data combination. Base means CMB priors + SNe Ia + BAO + BBN + CC data. Notice $\Delta\ln \mathcal{B}$ < 0 favors the $\Lambda$CDM}
	\end{figure}
	
	In order to compare the extended viscous dark energy with $\Lambda$CDM, we implement the Bayesian model comparison in terms of the strength of the evidence according to the Jeffreys scale. To do this, we estimate the values of the logarithm of the Bayesian evidence ($\ln \mathcal{E}$) and the Bayes factor ($\ln \mathcal{B}$). These values were obtained taking the priors and the dataset described in Sect. \ref{sec:data-method}. We assumed the $\Lambda$CDM model as the reference one. The Jeffreys scale interprets the Bayes factor as follows: inconclusive if $|\ln \mathcal{B}| < 1$, weak if $1 \leq |\ln \mathcal{B}| < 2.5$, moderate if $2.5 \leq |\ln \mathcal{B}| < 5$ and strong if $|\ln \mathcal{B}| \geq 5$. A negative (positive) value for $\ln\mathcal{B}$ indicates that the competing model is disfavored (supported) concerning the $\Lambda$CDM model.

	In Fig. \ref{fig:bayes-factor} we show the values obtained for the Bayes factor considering each model studied in this work. Looking at the Base data (red bars), we received that all models were disfavored by the data concerning the $\Lambda$CDM. We added the RSD data (blue bars); we got similar results. By considering the Base + $H_0$ local (green bars), we found that all viscous models were disfavored by the data ($\ln\mathcal{B} < 0$) but the q-$\Lambda$CDM received the positive value of the Bayes factor. Finally, taking into account the last data combination Base + RSD + $H_0$ local, we obtained the positive value for $\ln\mathcal{B}$ in the case of the q-$\Lambda$CDM model, and we achieved for Model 3 positive value for the Bayes factor.

	\subsection{Cosmic age crisis}
	
	The cosmic age crisis is an enduring issue in cosmology and is an important test to analyze the dynamical dark energy model. Because of several objects observed at intermediate and high redshifts, there are some challenges to accommodate a few of the old high redshift galaxies (OHRG) in the standard $\Lambda$CDM model \cite{Alcaniz1999,Alcaniz2003,Alcaniz2003.2,Jain2006}. Some approaches try to solve this problem in the literature. However, this issue has not solved, therefore, becoming a piece of important evidence for the favor of dark energy component \cite{Tammann:2001eg, Cui2010,Chen2012}. To check the age-consistency of the models studied in this work, we introduce the following ratio \cite{Alcaniz1999}
	
	\begin{equation}\label{eq:ratio-age}
		\tau(z_{\rm i}, \textbf{p}) = \frac{t(z_{\rm i}, \textbf{p})}{t_{\rm obs}(z_{\rm i})}, 
	\end{equation}
	where $t(z, \textbf{p})$ is the of Universe computed by
	
	\begin{equation}
		t(z, \textbf{p}) = \int_{z}^{\infty} \frac{dz'}{(1 + z')H(z, \textbf{p})},
	\end{equation}
	in which $H(z, \textbf{p})$ is given by Eqs. (\ref{eq:hubble1}), (\ref{eq:hubble2}) and (\ref{eq:hubble3}) and $t_{\rm obs}(z_{\rm i})$ is the observed age of the ith cosmological object. At a given redshift, the age of the Universe has to be larger than or compatible with the age of its oldest objects. Then we can translate this into $\tau \ge 1$, and in this way check the age-consistency of our models \cite{Alcaniz1999,Jain2006,Cui2010}.
	
	For this, we considered $32$ ages of passively evolving galaxies in the redshift range $0.117 \le z \le 1.845$ \cite{Simon2005}. Beside the quoted objects, we included a quasar APM $08279 +5255$ at $z = 3.91$ with an age $t = 2.1^{+0.9}_{-0.1}$ Gyr. We also check the age using nine extremely old globular clusters
	located in M31 \cite{Wang2010}.
	
	In Table \ref{tab:ages}, we report the present values of $\tau$ for extremely old globular clusters located in M31. By considering the data combination Base $+$ f$\sigma_8 + H_{0}$, we find that all models studied in this work cannot accommodate the object B050 with plausible statistical confidence. The other objects are accommodated by our models for at least $4\sigma$.
	
\begin{figure*}
		\centering
		\includegraphics[width=8.5cm]{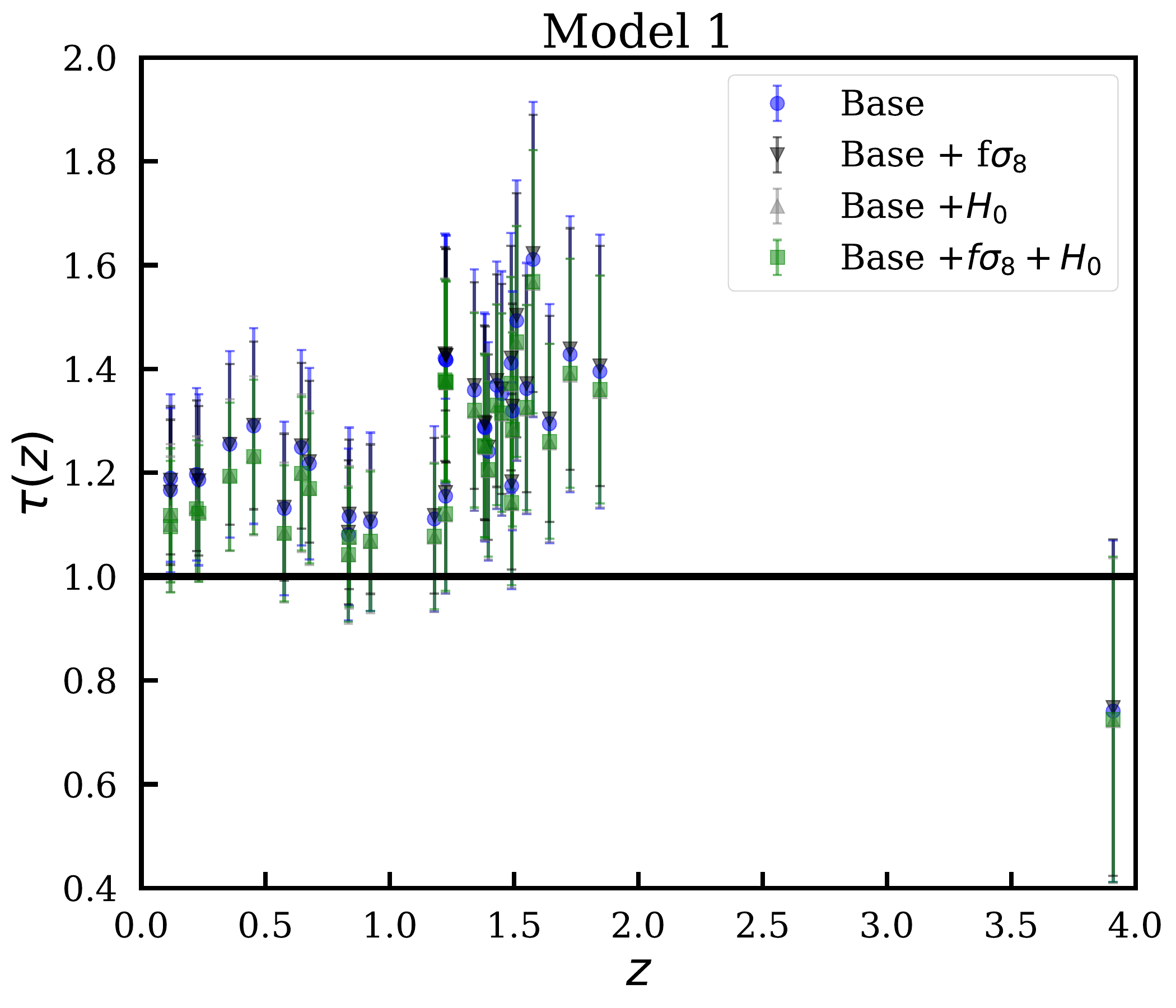}
		\includegraphics[width=8.5cm]{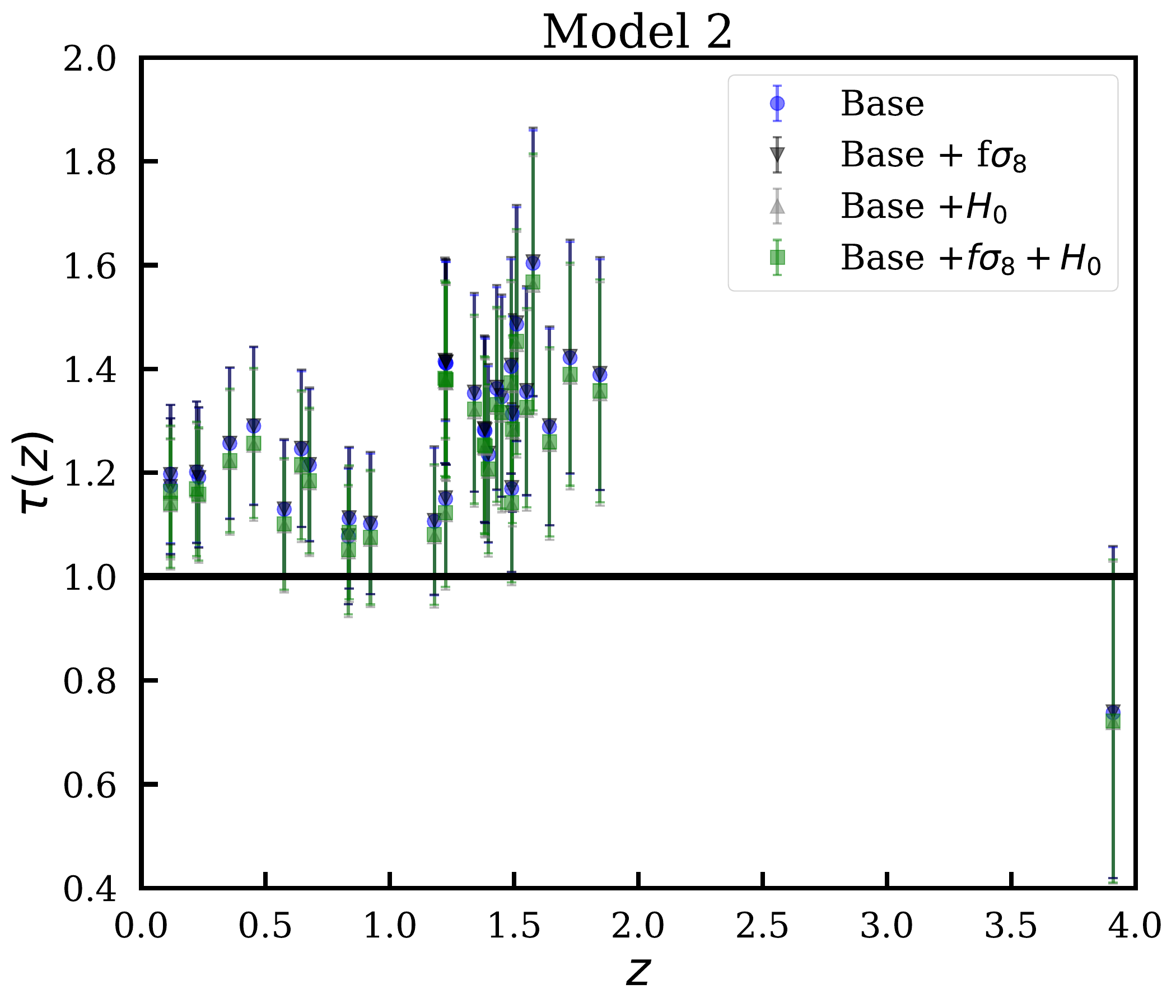}
		\includegraphics[width=8.5cm]{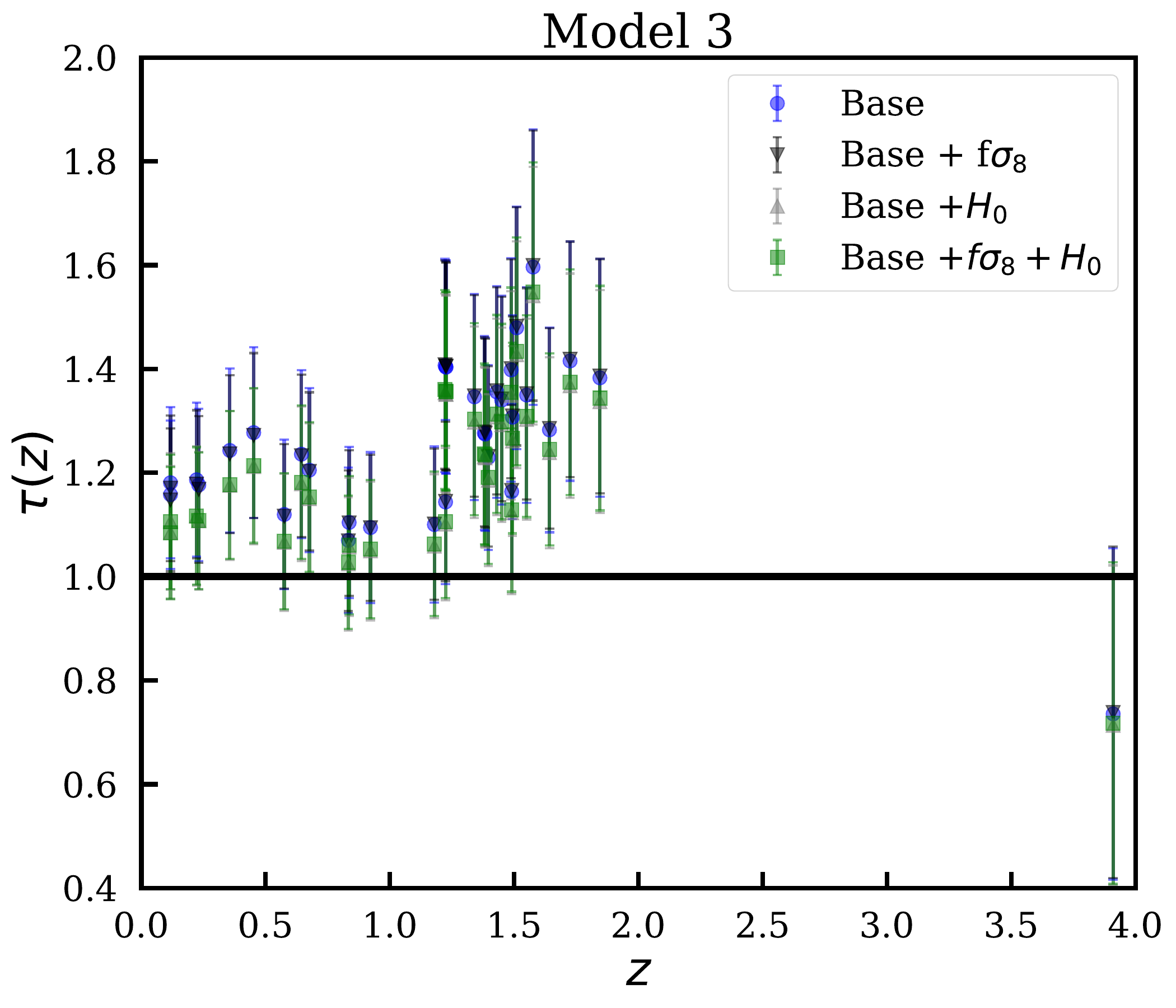}
		\includegraphics[width=8.5cm]{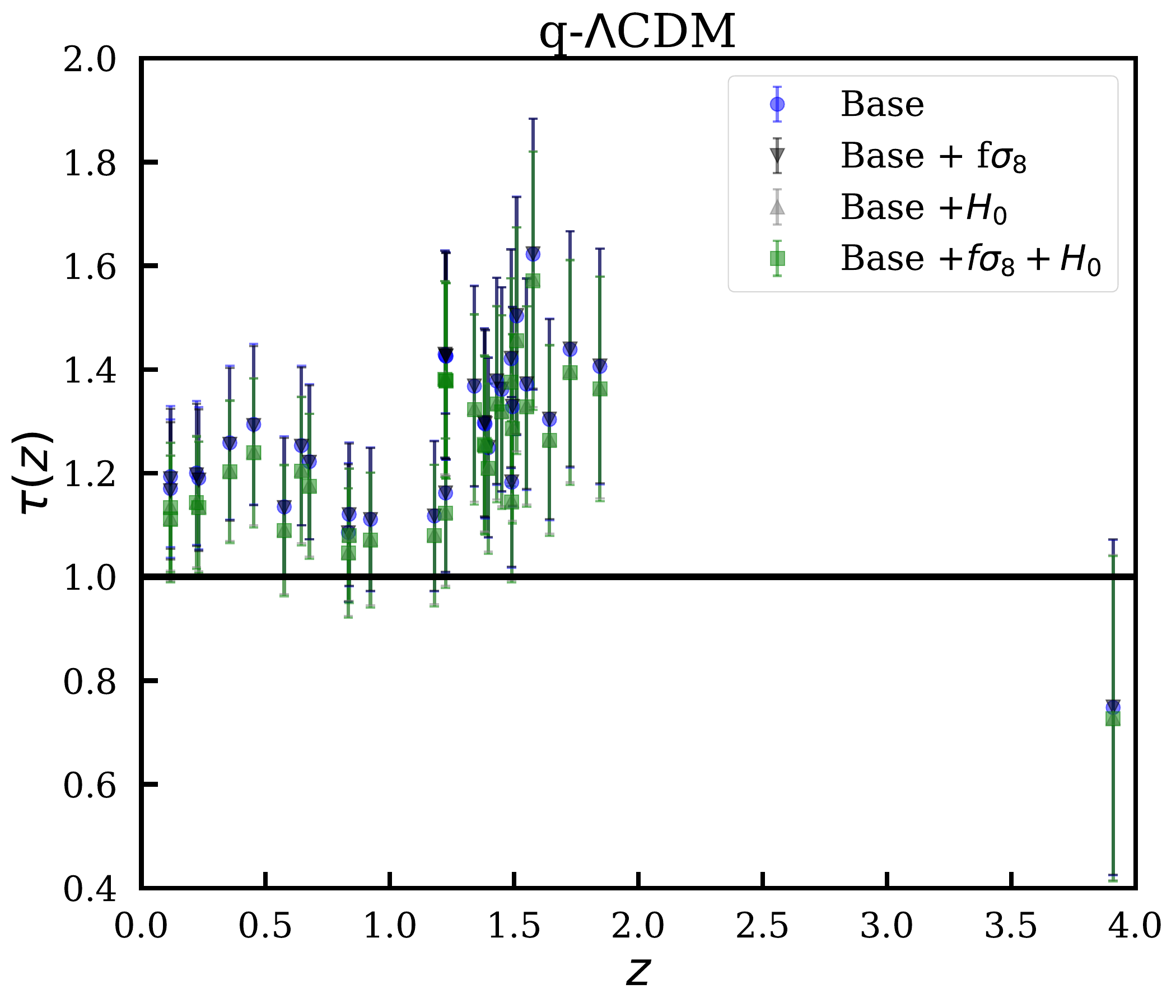}
		\caption{\label{fig:ages} The values of $\tau$ as function of redshift $z$ considering $32$ ages of passively evolving galaxies plus a age of quasar APM $08279 +5255$.}
	\end{figure*}

	Now, considering the ages of passively evolving galaxies, we show in Fig. \ref{fig:ages} the values of $\tau$ for all models assuming results obtained in the previous subsection. The results achieved were that models could accommodate the galaxies at least $1\sigma$ CL.
	
    \begin{table*}[t]
	\caption{\label{tab:ages} The values of $\tau$ considering $9$ extremely old globular clusters located in M31.}
	\begin{tabular}{lcccc}
    \hline
    Object & Model 1           & Model 2           & Model 3           & q-$\Lambda$CDM    \\ \hline
    B024   & $0.871 \pm 0.043$ & $0.890 \pm 0.056$ & $0.876 \pm 0.067$ & $0.876 \pm 0.058$ \\
    B050   & $0.831 \pm 0.016$ & $0.848 \pm 0.037$ & $0.835 \pm 0.051$ & $0.835 \pm 0.040$ \\
    B129   & $0.880 \pm 0.041$ & $0.898 \pm 0.055$ & $0.884 \pm 0.066$ & $0.885 \pm 0.057$ \\
    B144D  & $0.925 \pm 0.061$ & $0.945 \pm 0.073$ & $0.930 \pm 0.082$ & $0.931 \pm 0.074$ \\
    B239   & $0.916 \pm 0.130$ & $0.936 \pm 0.137$ & $0.920 \pm 0.141$ & $0.921 \pm 0.136$ \\
    B260   & $0.929 \pm 0.032$ & $0.949 \pm 0.050$ & $0.933 \pm 0.064$ & $0.934 \pm 0.053$ \\
    B297D  & $0.875 \pm 0.050$ & $0.894 \pm 0.061$ & $0.881 \pm 0.071$ & $0.879 \pm 0.063$ \\
    B383   & $0.950 \pm 0.071$ & $0.970 \pm 0.082$ & $0.955 \pm 0.091$ & $0.955 \pm 0.083$ \\
    B495   & $0.914 \pm 0.035$ & $0.934 \pm 0.050$ & $0.918 \pm 0.064$ & $0.919 \pm 0.053$ \\ \hline
    \end{tabular}
	\end{table*}

	\section{Conclusions}\label{sec:conclusions}

    In this paper, following the previous work on studying the viscous dark energy model inspired by Eckart's dissipative theory and the new interpretation of the Friedmann equations based on the Verlinde formalism \cite{Silva2019}. We investigated the linear perturbation theory of dark matter in the presence of a viscous dark energy model. 
    We implemented the relativistic perturbative equations for the extended viscous dark energy. Next, we investigated the matter perturbations' growth. The results obtained for the matter density contrast evolved similarly to the $\Lambda$CDM in high redshift when the extended viscous dark energy was subdominant. However, in late-time, the evolution of the contrast was slightly different from the standard model. We analyzed the influence of distinct parameter combinations on the linear growth rate evolution. Even though they evolved differently in intermediate redshift, we achieved that several models converged to a constant value. 
	
	By considering the extended viscous dark energy, we performed a Bayesian analysis using the latest background probes such as SNe Ia, BAO, cosmic chronometers, CMB priors, $H_0$ local, and the growth rate of perturbations from RSD data. We found that our viscous models were compatible with cosmological probes, and the $\Lambda$CDM model was recovered with $1\sigma$ CL. The models in this work relieve the tension of $H_0$ in $2 \sim 3 \sigma$. By addressing the $\sigma_8$ tension, our models can alleviate it. We compared the extended viscous dark energy with $\Lambda$CDM using the Bayesian evidence for all data combinations. This analysis demonstrated that both geometric data and growth data of structure discard extended viscous dark energy models. So, we could conclude that the holographic dark energy cannot fit the observations in cluster scales as good as the $\Lambda$CDM model. Finally, assuming passively evolving galaxies and high redshift objects, we analyzed our models in the cosmic age crisis framework and concluded that extended viscous dark energy models accommodate these objects.

	Future galaxy survey experiments such as J-PAS \cite{Benitez2014}, DESI \cite{Aghamousa2016}, and Euclid \cite{Euclid2011}, are expected to improve the quality of data and reduce the error bars in the determination of $f\sigma_8$ significantly, and thus will help to mitigate or exclude the tensions in cosmology.

	\begin{acknowledgements}
		The author thank CAPES and CNPq, Brazilian scientific support federal agencies, for financial support. RS thanks CNPq (Grant no. 303613/2015-7) for financial aid. This work was supported by the High-Performance Computing Center (NPAD)/UFRN.
	\end{acknowledgements}

	\bibliographystyle{apsrev4-2}
	\bibliography{references}

\end{document}